\documentclass{emulateapj}
\usepackage{apjfonts}

\newcommand{\Rsun}      {\mbox{$\rm\,R_{\mathord\odot}$}}

\begin{document}

\lefthead{{\em Chandra} Observations of {\em INTEGRAL} Sources}
\righthead{Tomsick et al.}

\submitted{To appear in ApJ}

\def\lsim{\mathrel{\lower .85ex\hbox{\rlap{$\sim$}\raise
.95ex\hbox{$<$} }}}
\def\gsim{\mathrel{\lower .80ex\hbox{\rlap{$\sim$}\raise
.90ex\hbox{$>$} }}}

\title{Localizing {\em INTEGRAL} Sources with {\em Chandra}:
X-Ray and Multi-Wavelength Identifications and Energy Spectra}

\author{John A. Tomsick\altaffilmark{1},
Arash Bodaghee\altaffilmark{1},
Sylvain Chaty\altaffilmark{2,3},
Jerome Rodriguez\altaffilmark{2},
Farid Rahoui\altaffilmark{4,5},
Jules Halpern\altaffilmark{6},
Emrah Kalemci\altaffilmark{7},
Mehtap \"Ozbey Arabac\i\altaffilmark{8}}

\altaffiltext{1}{Space Sciences Laboratory, 7 Gauss Way, 
University of California, Berkeley, CA 94720-7450, USA
(e-mail: jtomsick@ssl.berkeley.edu)}

\altaffiltext{2}{AIM (UMR-E 9005 CEA/DSM-CNRS-Universit\'e Paris Diderot) 
Irfu/Service d'Astrophysique, Centre de Saclay, FR-91191 Gif-sur-Yvette Cedex, 
France}

\altaffiltext{3}{Institut Universitaire de France, 103, bd Saint-Michel, 
75005 Paris, France}

\altaffiltext{4}{Harvard University, Astronomy Department, 60 Garden Street, 
Cambridge, MA 02138, USA}

\altaffiltext{5}{Harvard-Smithsonian Center for Astrophysics, 60 Garden Street,
Cambridge, MA 02138, USA}

\altaffiltext{6}{Columbia Astrophysics Laboratory, Columbia University, 
550 West 120th Street, New York, NY 10027-6601, USA}

\altaffiltext{7}{Sabanc\i\ University, Orhanl\i -Tuzla, \.Istanbul, 34956, Turkey}

\altaffiltext{8}{Physics Department, Middle East Technical University, Ankara, 
06531, Turkey}

\begin{abstract}

We report on {\em Chandra} observations of 18 hard X-ray ($>$20\,keV) sources 
discovered with the {\em INTEGRAL} satellite near the Galactic plane.  For 14
of the {\em INTEGRAL} sources, we have uncovered one or two potential {\em Chandra} 
counterparts per source.  These provide soft X-ray (0.3--10\,keV) spectra and 
sub-arcsecond localizations, which we use to identify counterparts at other 
wavelengths, providing information about the nature of each source.  Despite the 
fact that all of the sources are within 5$^{\circ}$ of the plane, four of the IGR 
sources are AGN (IGR~J01545+6437, IGR~J15391--5307, IGR~J15415--5029, and 
IGR~J21565+5948) and four others are likely AGN (IGR~J03103+5706, IGR~J09189--4418, 
IGR~J16413--4046, and IGR~J16560--4958) based on each of them having a strong IR 
excess and/or extended optical or near-IR emission.  We compare the X-ray and 
near-IR fluxes of this group of sources to those of AGN selected by their 2--10\,keV 
emission in previous studies and find that these IGR AGN are in the range of typical 
values.  There is evidence in favor of four of the sources being Galactic 
(IGR~J12489--6243, IGR~J15293--5609, IGR~J16173--5023, and IGR~J16206--5253), but 
only IGR~J15293--5609 is confirmed as a Galactic source as it has a unique 
{\em Chandra} counterpart and a parallax measurement from previous optical 
observations that puts its distance at $1.56\pm 0.12$\,kpc.  The 0.3--10\,keV 
luminosity for this source is $(1.4^{+1.0}_{-0.4})\times 10^{32}$\,erg\,s$^{-1}$, and 
its optical/IR spectral energy distribution is well described by a blackbody with a 
temperature of 4200--7000\,K and a radius of 12.0--16.4\,\Rsun.  These values suggest 
that IGR~J15293--5609 is a symbiotic binary with an early K-type giant and a white 
dwarf accretor.  We also obtained likely {\em Chandra} identifications for 
IGR~J13402--6428 and IGR~J15368--5102, but follow-up observations are required to 
constrain their source types.

\end{abstract}

\keywords{galaxies: active --- stars: white dwarfs --- stars: neutron ---
X-rays: galaxies --- X-rays: stars ---
stars: individual (IGR~J01545+6437, IGR~J03103+5706, IGR~J04069+5042,
IGR~J06552--1146, IGR~J09189--4418, IGR~J12489--6243, IGR~J13402--6428, 
IGR~J15293--5609, IGR~J15368--5102, IGR~J15391--5307, IGR~J15415--5029, 
IGR~J16173--5023, IGR~J16206--5253, IGR~J16413--4046, IGR~J16560--4958, 
IGR~J21188+4901,  IGR~J21565+5948,  IGR~J22014+6034)}

\section{Introduction}

The hard X-ray imaging of the Galactic plane by the {\em INTErnational 
Gamma-Ray Astrophysics Laboratory (INTEGRAL)} satellite continues to
find new ``IGR'' sources at energies $>$20\,keV.  While {\em INTEGRAL} 
excels at discovering new sources in the 20--40\,keV band, especially due 
to its large field-of-view, it only localizes them to $1^{\prime}$--$5^{\prime}$, 
which is not nearly adequate for finding optical/IR counterparts.  Short 
{\em Chandra} exposures of IGR sources allow for a major advance in 
understanding the nature of these sources by providing sub-arcsecond 
positions, leading to unique optical/IR counterparts, as well as 0.3--10\,keV 
spectra that can be used to measure column densities and continuum shapes.  

After observations between 2003 and 2008, {\em INTEGRAL} had detected 
more than 700 hard X-ray sources \citep{bird10}.  While that catalog 
includes nearly 400 IGR sources, the most up-to-date 
list\footnote{See http://irfu.cea.fr/Sap/IGR-Sources/.} includes more 
than 500 IGR entries.  The largest group of identified sources are the
Active Galactic Nuclei with $>$250 confirmed AGN detected \citep{bird10}.
The largest Galactic source types include High-Mass X-ray Binaries (HMXBs), 
Low-Mass X-ray Binaries (LMXBs), and Cataclysmic Variables (CVs).  A subset 
of these binaries are symbiotics with a white dwarf or neutron star 
accreting from a giant (i.e., luminosity class III) star.  It has been
somewhat surprising that the symbiotics with white dwarf accretors 
can produce hard X-ray emission \citep{kennea09,ratti10}.  In addition, 
{\em INTEGRAL} has detected new as well as known supernova remnants (SNRs) 
and pulsar wind nebulae (PWNe).  Unidentified IGR sources is also still a 
large group, and the most critical piece of information for determining the 
nature of these sources is to obtain a sub-arcsecond X-ray position.  

\begin{table*}
\begin{center}
\caption{{\em Chandra} Observations\label{tab:obs}}
\begin{tabular}{cccclc} \hline \hline
IGR Name & ObsID & $l$\footnote{Galactic longitude in degrees measured by {\em INTEGRAL}.} & $b$\footnote{Galactic latitude in degrees measured by {\em INTEGRAL}.} & Start Time (UT) & Exposure (s)\\ \hline \hline
J01545+6437   & 12425 & 129.62 &  +2.57 & 2011 Feb  7, 19.47 h & 4987\\
J03103+5706   & 12419 & 140.95 & --0.83 & 2010 Nov 23, 10.39 h & 4716\\
J04069+5042   & 12421 & 151.43 & --1.03 & 2010 Nov 25,  5.47 h & 4987\\
J06552--1146  & 12433 & 223.85 & --4.52 & 2010 Dec 12, 22.35 h & 4986\\
J09189--4418  & 12426 & 267.93 &  +3.64 & 2011 Jul 22,  0.37 h & 5007\\
J12489--6243  & 12416 & 302.64 &  +0.15 & 2010 Nov 25, 22.19 h & 4986\\
J13402--6428  & 12424 & 308.15 & --2.10 & 2011 Oct  6,  9.38 h & 4995\\
J15293--5609  & 12417 & 323.66 &  +0.22 & 2011 Jun 22, 22.08 h & 5096\\
J15368--5102  & 12428 & 327.52 &  +3.76 & 2011 Oct  9,  2.91 h & 4695\\
J15391--5307  & 12422 & 326.58 &  +1.88 & 2011 Apr 14, 21.18 h & 4992\\
J15415--5029  & 12429 & 328.45 &  +3.76 & 2011 Oct  9,  1.13 h & 4994\\
J16173--5023  & 12415 & 332.84 &  +0.14 & 2011 Oct 11,  5.28 h & 4989\\
J16206--5253  & 12423 & 331.47 & --2.01 & 2011 Jul  5, 18.10 h & 4992\\
J16413--4046  & 12427 & 342.71 &  +3.70 & 2011 May  9,  0.50 h & 4991\\
J16560--4958  & 12431 & 337.32 & --4.17 & 2011 May 23,  5.91 h & 4992\\
J21188+4901   & 12418 &  91.27 & --0.33 & 2011 Feb 28,  2.77 h & 4989\\
J21565+5948   & 12430 & 102.54 &  +4.06 & 2011 May 30, 20.79 h & 4992\\
J22014+6034   & 12432 & 103.49 &  +4.28 & 2011 Apr 16, 11.82 h & 4994\\ \hline
\end{tabular}
\end{center}
\end{table*}

Even the faintest {\em INTEGRAL} detections, which are near 
$\sim$$2\times 10^{-12}$\,erg\,cm$^{-2}$\,s$^{-1}$ (20--40\,keV), are 
usually easily detected by {\em Chandra} in a short exposure time.  With 
our previous {\em Chandra} programs, we obtained 5\,ks snapshots of 46 IGR 
sources, and obtained 35 {\em Chandra} counterparts 
\citep{tomsick06,tomsick08a,tomsick09a}.  Based on {\em Chandra} spectral 
information and optical/IR observations 
\citep[e.g.,][]{masetti7,chaty08,butler09}, definite or likely 
identifications have been obtained for 32 of our {\em Chandra} targets.  
These programs have all concentrated on low Galactic latitude targets
as the {\em Chandra} positions are most critical for obtaining optical
and IR counterparts in the crowded regions of the Galactic plane.  
Although our scientific goals have leaned toward finding Galactic
sources such as obscured HMXBs \citep{mg03,fc04,walter06,cr12} or supergiant 
fast X-ray transients \citep[SFXTs,][]{negueruela06,smith06,pellizza06}, 
it has still been relatively common to find that the {\em INTEGRAL} sources 
are background AGN \citep[e.g.,][]{zct09}.

In the following, we describe the results of {\em Chandra} observations
of 18 IGR sources that were made between late-2010 and late-2011.
The observations are described in \S\,2, and the search for X-ray 
counterparts, fits to the {\em Chandra} energy spectra, and multi-wavelength
identifications are described in \S\,3.  A discussion of the results for
each source is provided in \S\,4, and \S\,5 includes a summary.

\section{{\em Chandra} Observations and Data Processing}

When the target selection was done for this program, we considered 723 
sources listed in the 4$^{\mathrm{th}}$ {\em INTEGRAL} survey catalog of 
\cite{bird10}. We excluded sources if their classification (e.g. AGN, 
HMXB, etc.) was known according to \cite{bird10} or the {\em INTEGRAL}
Sources website$^{1}$, and we only considered sources situated within 
5$^{\circ}$ of the Galactic plane.  From 98 low-latitude sources, we 
removed sources labeled as transient in \cite{bird10}, as well as sources 
with previous or planned observations by {\em Chandra}, {\em XMM-Newton}, 
or {\em Swift}. 

After the selections described above, we were left with 19 sources 
that we observed with {\em Chandra}.  We reported on one target,
IGR~J11014--6103, in a separate paper \citep{tomsick12}
and the other 18 targets here.  Table~\ref{tab:obs} lists the sources, 
their {\em INTEGRAL} positions, and basic information about the 
{\em Chandra} observations.  The observations of the IGR source fields 
occurred between 2010 November 23 and 2011 October 11, and, in each case, 
an exposure time of $\sim$5~ks was obtained.  We used the  Advanced CCD 
Imaging Spectrometer \citep[ACIS,][]{garmire03} instrument, which has
a 0.3--10\,keV bandpass.  The 90\% confidence {\em INTEGRAL} error circles 
range from $3.\!^{\prime}3$ to $5.\!^{\prime}4$, leading us to use the 
$16.9\times 16.9$\,arcmin$^{2}$ field-of-view (FOV) of the ACIS-I 
instrument.  We used the ACIS ``Faint'' mode because of the potential
for bright sources to saturate the {\em Chandra} telemetry if we used
``Very Faint'' mode.  In addition, we used a custom dither pattern 
to decrease the non-uniformity of the exposure time due to gaps between
the four ACIS-I chips.

The data were initially processed at the {\em Chandra} X-ray Center (CXC) 
with ASCDS versions between 8.3.3.1 and 8.4.  After obtaining the data from 
the CXC, we performed all subsequent processing with the {\em Chandra} 
Interactive Analysis of Observations (CIAO) version 4.3.1 software and 
Calibration Data Base (CALDB) version 4.4.6.  We used the CIAO program 
{\ttfamily chandra\_repro} to produce the ``level 2'' event lists that we 
used for further analysis.

\begin{table*}
\caption{{\em Chandra} Candidate Counterparts to IGR Sources\label{tab:counterparts}}
\begin{minipage}{\linewidth}
\begin{center}
\begin{tabular}{llccccc} \hline \hline
 & & $\theta$\footnote{The angular distance between the center of the {\em INTEGRAL} error circle, which is also the approximate {\em Chandra} aimpoint, and the source.}/$\theta_{INTEGRAL}$\footnote{The size of the 90\% confidence {\em INTEGRAL} error radius given in \cite{bird10}.} & {\em Chandra} R.A.\footnote{The position uncertainties for these relatively bright sources are dominated by the systematic error, which is $0.\!^{\prime\prime}64$ at 90\% confidence and $1^{\prime\prime}$ at 99\% confidence \citep{weisskopf05}.} & {\em Chandra} Decl.$^{\it c}$ & ACIS & \\
IGR Name & CXOU Name & (arcminutes) & (J2000)  & (J2000) &  Counts\footnote{The number of counts, after background subtraction, measured by {\em Chandra}/ACIS-I in the 0.3--10\,keV band.} & Hardness\footnote{The hardness is given by $(C_{2}-C_{1})/(C_{2}+C_{1})$, where $C_{2}$ is the number of counts in the 2--10 keV band and $C_{1}$ is the number of counts in the 0.3--2 keV band.}\\ \hline \hline
J01545+6437 & J015435.2+643757 (1)  &  1.32/4.1 & $01^{\rm h}54^{\rm m}35^{\rm s}.28$ &  +$64^{\circ}37^{\prime}57.\!^{\prime\prime}8$ &  29.3 &  +$0.40\pm 0.26$\\ \hline 
J03103+5706 & J031010.6+570712 (2)  &  1.37/4.9 & $03^{\rm h}10^{\rm m}10^{\rm s}.61$ &  +$57^{\circ}07^{\prime}12.\!^{\prime\prime}1$ &  17.4 &  +$0.79\pm 0.41$\\ \hline 
J09189--4418 & J091858.8--441830 (3) &  0.64/4.5 & $09^{\rm h}18^{\rm m}58^{\rm s}.83$ & --$44^{\circ}18^{\prime}30.\!^{\prime\prime}9$ & 377.4 &  +$0.79\pm 0.07$\\ \hline 
J12489--6243 & J124846.4--623743 (4a) &  5.42/4.1 & $12^{\rm h}48^{\rm m}46^{\rm s}.44$ & --$62^{\circ}37^{\prime}43.\!^{\prime\prime}1$ &  38.1 &  +$0.67\pm 0.24$\\ 
 & J124905.3--624242 (4b) &  1.40/4.1 & $12^{\rm h}49^{\rm m}05^{\rm s}.32$ & --$62^{\circ}42^{\prime}42.\!^{\prime\prime}6$ &  16.3 &  +$0.16\pm 0.35$\\ \hline 
J13402--6428 & J133935.8--642537 (5a) &  5.02/3.5 & $13^{\rm h}39^{\rm m}35^{\rm s}.89$ & --$64^{\circ}25^{\prime}37.\!^{\prime\prime}7$ &  97.7 &  +$0.38\pm 0.13$\\ 
 & J133959.2--642444 (5b) &  4.29/3.5 & $13^{\rm h}39^{\rm m}59^{\rm s}.25$ & --$64^{\circ}24^{\prime}44.\!^{\prime\prime}3$ &  31.7 &  +$0.16\pm 0.24$\\ \hline 
J15293--5609 & J152929.3--561213 (6) &  3.08/5.4 & $15^{\rm h}29^{\rm m}29^{\rm s}.37$ & --$56^{\circ}12^{\prime}13.\!^{\prime\prime}3$ & 119.4 & --$0.51\pm 0.12$\\ \hline 
J15368--5102 & J153658.6--510221 (7) &  1.82/4.9 & $15^{\rm h}36^{\rm m}58^{\rm s}.64$ & --$51^{\circ}02^{\prime}21.\!^{\prime\prime}4$ &  77.4 &  +$0.23\pm 0.14$\\ \hline 
J15391--5307 & J153916.7--530815 (8) &  2.25/4.5 & $15^{\rm h}39^{\rm m}16^{\rm s}.78$ & --$53^{\circ}08^{\prime}15.\!^{\prime\prime}9$ &  46.5 &  +$1.01\pm 0.25$\\ \hline 
J15415--5029 & J154126.4--502823 (9a) &  1.12/3.8 & $15^{\rm h}41^{\rm m}26^{\rm s}.45$ & --$50^{\circ}28^{\prime}23.\!^{\prime\prime}5$ &  46.4 &  +$0.71\pm 0.22$\\ 
 & J154140.5--502848 (9b) &  2.02/3.8 & $15^{\rm h}41^{\rm m}40^{\rm s}.57$ & --$50^{\circ}28^{\prime}48.\!^{\prime\prime}6$ &  15.4 &  +$0.11\pm 0.37$\\ \hline 
J16173--5023 & J161728.2--502242 (10a) &  2.11/3.3 & $16^{\rm h}17^{\rm m}28^{\rm s}.26$ & --$50^{\circ}22^{\prime}42.\!^{\prime\prime}5$ & 694.6 &  +$0.36\pm 0.04$\\ 
 & J161720.6--502415 (10b) &  1.38/3.3 & $16^{\rm h}17^{\rm m}20^{\rm s}.65$ & --$50^{\circ}24^{\prime}15.\!^{\prime\prime}1$ &  15.4 &  +$1.02\pm 0.49$\\ \hline 
J16206--5253 & J161955.4--525230 (11) &  6.09/4.6 & $16^{\rm h}19^{\rm m}55^{\rm s}.49$ & --$52^{\circ}52^{\prime}30.\!^{\prime\prime}2$ &  70.3 &  +$0.80\pm 0.18$\\ \hline 
J16413--4046 & J164119.4--404737 (12) &  0.81/4.5 & $16^{\rm h}41^{\rm m}19^{\rm s}.49$ & --$40^{\circ}47^{\prime}37.\!^{\prime\prime}7$ & 147.4 &  +$0.91\pm 0.12$\\ \hline 
J16560--4958 & J165551.9--495732 (13) &  0.99/4.1 & $16^{\rm h}55^{\rm m}51^{\rm s}.95$ & --$49^{\circ}57^{\prime}32.\!^{\prime\prime}4$ & 1092.4 & +$0.54\pm 0.04$\\ \hline 
J21565+5948 & J215604.2+595604 (14) &  7.88/4.9 & $21^{\rm h}56^{\rm m}04^{\rm s}.22$ &  +$59^{\circ}56^{\prime}04.\!^{\prime\prime}1$ & 202.8 &  +$0.11\pm 0.08$\\ \hline 
\end{tabular}
\end{center}
\end{minipage}
\end{table*}

\section{Analysis and General Results}

\subsection{Search for X-ray Counterparts to the IGR Sources}

We used the CIAO routine {\ttfamily wavdetect} to search for X-ray sources 
on the ACIS-I chips.  For each observation, we searched for sources in 
unbinned images with $2048\times 2048$ pixels as well as images binned by 
a factor of 2 ($1024\times 1024$ pixels), a factor of 4 ($512\times 512$ 
pixels), and a factor of 8 ($256\times 256$ pixels).  In each case, we set 
the detection threshold to a level that would be expected to yield one 
spurious source ($2.4\times 10^{-7}$, $9.5\times 10^{-7}$, $3.8\times 10^{-6}$, 
and $1.5\times 10^{-5}$, respectively, corresponding to the inverse of the 
number of pixels for each of the four images).  Thus, in the merged lists 
of detected sources, it would not be surprising if a few of the sources in 
each field are spurious.  

Although we typically detected dozens of {\em Chandra} sources in each field, 
we used information about their brightnesses, their hardnesses, and their 
locations relative to the {\em INTEGRAL} error circle to determine which 
sources to consider as candidate counterparts to the {\em INTEGRAL} sources.
To determine the most likely candidates, we followed the technique used in 
\cite{tomsick09a}.  The broadest Galactic plane survey to date that resulted 
in a suitable $\log{N}$-$\log{S}$ curve was obtained with the {\em Advanced 
Satellite for Cosmology and Astrophysics (ASCA)}.  The 2--10\,keV 
$\log{N}$-$\log{S}$ is described as a power-law with 
$N(>F_{2-10~\rm keV}) = 9.2(F_{2-10~\rm keV}/10^{-13})^{-0.79}$\,deg$^{-2}$, where 
$F_{2-10~\rm keV}$ is the absorbed 2--10\,keV flux in units of erg\,cm$^{-2}$\,s$^{-1}$
\citep{sugizaki01}.  Although this relationship is meant to be for Galactic 
sources only, \cite{sugizaki01} as well as \cite{hands04}, which combines the 
{\em ASCA} measurements with {\em Chandra} and {\em XMM-Newton}
$\log{N}$-$\log{S}$ measurements for smaller regions of the plane, suggest 
that the above relationship is a good approximation to the total source 
density down to a flux level of a few $\times 10^{-13}$ erg\,cm$^{-2}$\,s$^{-1}$.  
Thus, we use this relationship for our initial calculations, but we discuss 
the impact of a steeper $\log{N}$-$\log{S}$ on our faintest {\em Chandra} 
candidates in \S\,3.3.  Using Poisson statistics, the probability ($P$) that 
a source is spurious is given by 
\begin{equation}
P = 1 - e^{-N(>F_{2-10~\rm keV})~\pi~\theta_{\rm search}^{2}}~~~,
\end{equation}
where $\theta_{\rm search}$ is the radius of the search region in units of degrees.
In cases where a source is within the 90\% confidence {\em INTEGRAL} error circle, 
which has a radius of $\theta_{INTEGRAL}$, the error circle is the search region
(i.e., $\theta_{\rm search} = \theta_{INTEGRAL}$).  For sources outside the error circle, 
$\theta_{\rm search} = \theta$, where $\theta$ is the angular separation between the 
source and the center of the {\em INTEGRAL} error circle.

While calculating the absolute spurious probabilities requires spectral modeling 
to determine source fluxes, here, we calculate a relative probability, $P_{\rm rel}$, 
using the 2--10\,keV count rates, $C_{2-10~\rm keV}$ according to
\begin{equation}
P_{\rm rel} = 1 - e^{-(\frac{C_{2-10\,\rm keV}}{C_{0}})^{-0.79}~\pi~\theta_{\rm search}^{2}}~~~,
\end{equation}
where $C_{0}$ is an arbitrary number of counts.  We set this parameter to 520, 
which makes the source with the highest number of counts in the 2-10\,keV
band have $P_{\rm rel} = 0.01$.

After carrying out the spectral analysis (\S\,3.2) and determining the 
actual fluxes for a subset of the sources, we find that a value of 
$P_{\rm rel} = 0.3$ corresponds to a 10--25\% spurious probability, where
the large range is caused by the variation in the counts-to-flux ratio 
with the hardness of the spectrum.  We consider $P_{\rm rel} = 0.3$ to be 
our threshold for further consideration of a source as a candidate 
counterpart.  We made only two exceptions to the strict threshold: for 
IGR~J09189--4418, we do not consider two candidates with $P_{\rm rel} = 0.29$ 
because there is a much better candidate with $P_{\rm rel} = 0.02$, and for 
IGR~J16413--4046, we do not consider a candidate that is right at the 
threshold because there is a candidate with $P_{\rm rel} = 0.05$.

With these criteria, we find four cases with no {\em Chandra} sources 
that stand out from the general X-ray source population: IGR~J04069+5042, 
IGR~J06552--1146, IGR~J21188+4901, and IGR~J22014+6034, and for these 
sources, we include lists of detected sources in these fields in Appendix A.  
Ten sources have unique candidate counterparts, and four sources have two 
candidate counterparts.  The final list of 18 sources that we consider as 
potential or likely counterparts to the IGR sources is given in 
Table~\ref{tab:counterparts}.  For IGR sources with two {\em Chandra} 
counterparts, the first source is the one with lower $P_{\rm rel}$.  

\begin{figure}
\includegraphics[scale=0.5]{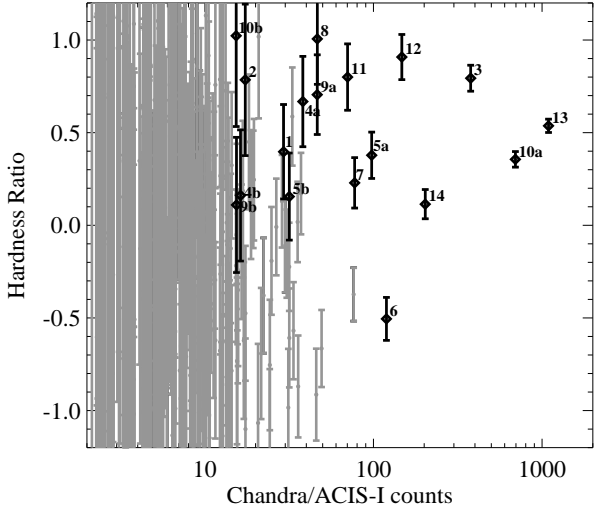}
\caption{Hardness-intensity diagram, including all of the {\em Chandra}/ACIS-I sources detected in
the 18 {\em Chandra} observations.  The intensity is given as the number of ACIS-I counts in the 
0.3--10~keV energy band.  The hardness is given by $(C_{2}-C_{1})/(C_{2}+C_{1})$, where $C_{2}$ is
the number of counts in the 2--10~keV band and $C_{1}$ is the number of counts in the 0.3--2~keV
band.  The sources that we consider as possible {\em Chandra} counterparts to the IGR sources are 
plotted with black diamonds and error bars.  The labels correspond to the sources as follows: 
1=IGR~J01545+6437, 2=IGR~J03103+5706, 3=IGR~J09189--4418, 4a=IGR~J12489--6243 (candidate a), 
4b=IGR~J12489--6243 (candidate b), 5a=IGR~J13402--6428 (candidate a), 5b=IGR~J13402--6428 
(candidate b), 6=IGR~J15293--5609, 7=IGR~J15368--5102, 8=IGR~J15391--5307, 
9a=IGR~J15415--5029 (candidate a), 9b=IGR~J15415--5029 (candidate b), 
10a=IGR~J16173--5023 (candidate a), 10b=IGR~J16173--5023 (candidate b), 11=IGR~J16206--5253,
12=IGR~J16413--4046, 13=IGR~J16560--4958, and 14=IGR~J21565+5948.
\label{fig:hi}}
\end{figure}

The hardness-intensity diagram, including all of the detected {\em Chandra} 
sources, is shown in Figure~\ref{fig:hi}.  For each source, we determined
the number of source counts in the 0.3--2\,keV band ($C_{1}$) and the 2--10\,keV 
band ($C_{2}$).  Using the size of the {\em Chandra} point spread function 
(PSF) as a guide, we used extraction radii of $5^{\prime\prime}$ for sources 
within $4^{\prime}$ of the aimpoint, $10^{\prime\prime}$ for sources between 
$4^{\prime}$ and $7^{\prime}$ from the aimpoint, and $15^{\prime\prime}$ for 
sources more than $7^{\prime}$ from the aimpoint.  The background counts were 
taken from a rectangular source-free region, scaled to the size of the source 
extraction region, and subtracted off.  We then calculated the hardness 
according to $(C_{2}-C_{1})/(C_{2}+C_{1})$, which runs from --1.0 for the 
softest sources (all the counts in the 0.3--2\,keV band) to +1.0 for the 
hardest sources (all the counts in the 2--10\,keV band).  To deal with the
case where there are zero counts in one of the energy bands, we used the 
``Gehrels'' prescription for determining the uncertainties \citep{gehrels86}.
The sources that we are considering as candidate counterparts are labeled, 
and they primarily occupy the region of the diagram with harder and brighter 
sources.  For these sources, the numbers of source counts and the hardnesses
are given in Table~\ref{tab:counterparts}.

\subsection{Energy Spectra of the Candidate Counterparts}

For each of the candidate counterparts, we produced and modeled the 
0.3--10\,keV {\em Chandra} energy spectrum.  We extracted source photons
from circular regions with radii as described above.  Although background
is negligible in most cases, we made a background estimate using the 
rectangular regions described above, and performed background subtraction.  
We fitted the spectra using the XSPEC v12.7.0 software package.  For 
16 of the 18 spectra, we fitted unbinned spectra and optimized the fit 
using Cash statistics\footnote{Although we follow the XSPEC terminology 
by calling this the ``Cash'' statistic, when background subtraction is
performed, it is technically the W-statistic that is minimized.
See http://heasarc.nasa.gov/xanadu/xspec/manual/XSappendixStatistics.html.}.  
In two cases, the count rates were high enough to cause photon pile-up, 
and we used $\chi^{2}$ statistics in these cases.  We accounted for 
pile-up in the two cases using the XSPEC model {\ttfamily pileup} 
\citep{davis01}.

We fitted the spectra with a model consisting of an absorbed power-law
and used \cite{wam00} abundances and \cite{bm92} cross sections to model
the absorption, and the results are shown in Table~\ref{tab:spectra}.
For each source, the measured column density potentially includes 
contributions from the Galactic ISM as well as from absorption that is
intrinsic to the source.  The value of $N_{\rm H}$ given in 
Table~\ref{tab:spectra} is the total, but we also include the 
Galactic $N_{\rm H}$ from \cite{kalberla05} for comparison.
For the spectra fitted with Cash statistics, we used the XSPEC routine 
{\ttfamily goodness} to determine the percentage of simulated spectra 
with a fit statistic less than that obtained for the data.  One expects 
a value near 50\% if the model is consistent with the spectrum, and 
we obtain an average value of 55\%.  The values range from 28.6\% to 
81.4\% with three sources having values $>$75\%.  For these three
cases, we produced binned spectra with at least 10 counts per bin, 
performed a $\chi^{2}$ minimization fit, and inspected the residuals.  
For CXOU~J091858.8--441830 and CXOU~J215604.2+595604, we obtained 
reduced-$\chi^{2}$ ($\chi^{2}_{\nu}$) values of 1.0 (33 degrees of freedom) 
and 0.85 (17 dof), respectively, and there are no obvious features in 
the residuals in either case.  For CXOU~J164119.4--404737, 
$\chi^{2}_{\nu} = 1.2$ (12 dof), and there are positive residuals near 5~keV.  
Adding a Gaussian with an energy of $5.0\pm 0.3$\,keV provides a marginally
significant improvement in the fit to $\chi^{2}_{\nu} = 0.95$ (9 dof).  It is 
possible that the source is a background AGN and that the Gaussian is a
redshifted iron line.  The line energy implies a redshift of 
$z = 0.28\pm 0.08$.  As the detection of this emission line is marginal, 
we report the fit parameters with and without the line in Table~\ref{tab:spectra}.  

\begin{table*}
\caption{{\em Chandra} Spectral Results\label{tab:spectra}}
\begin{minipage}{\linewidth}
\begin{center}
\begin{tabular}{llcccccc} \hline \hline
 & & $N_{\rm H}$\footnote{The parameters are for power-law fits to the {\em Chandra}/ACIS spectra and include photoelectric absorption with cross sections from \cite{bm92} and abundances from \cite{wam00}.  In general, the measured value of $N_{\rm H}$ and the errors on $N_{\rm H}$ are scaled-down by $\sim$30\% if solar abundances \citep{ag89} are used rather than the approximation to average interstellar abundances \citep{wam00}.  A pile-up correction was applied in the two cases where pile-up parameters are given.  The PSF fraction \citep{davis01} was left fixed to 0.95 (the default value).  Errors in this table are at the 90\% confidence level ($\Delta$$C = 2.7$ for the Cash fits and $\Delta$$\chi^{2} = 2.7$ for the $\chi^{2}$ fits).} 

&     & X-ray &  & Fit & Galactic $N_{\rm H}$\\
IGR Name & CXOU Name & ($\times 10^{22}$ cm$^{-2}$) & $\Gamma$ & Flux\footnote{Unabsorbed 0.3--10 keV flux in units of $10^{-12}$ erg~cm$^{-2}$~s$^{-1}$.} & $\alpha$\footnote{The grade migration parameter in the pile-up model \citep{davis01}.  The probability that $n$ events will be piled together but will still be retained after data filtering is $\alpha^{n-1}$.} & Statistic\footnote{For the Cash fits, we report the Cash statistic for the best fit model, the degrees of freedom, and, in parentheses, the percentage of simulated spectra for which better fits were obtained.  For the $\chi^{2}$ fits, we report the $\chi^{2}$ for the best fit model and the degrees of freedom.} & ($\times 10^{22}$ cm$^{-2}$)\footnote{The column density inferred from the \cite{kalberla05} H\,I survey.}\\ \hline \hline
J01545+6437 & J015435.2+643757 (1)  & $0.0^{+0.4}_{-0.0}$  & --$0.40^{+0.56}_{-0.59}$ & $0.30^{+0.16}_{-0.11}$ & --- & 177.4/660 (54.6\%) & 0.67\\ \hline
J03103+5706 & J031010.6+570712 (2)  & $4.5^{+15.5}_{-3.7}$ &   $2.2^{+5.1}_{-1.8}$    & $0.9^{+90,000}_{-0.6}$ & --- & 109.7/660 (43.2\%) & 0.74\\ \hline
J09189--4418 & J091858.8--441830 (3) & $4.6^{+1.1}_{-1.3}$ &    $1.4^{+0.4}_{-0.4}$ & $4.1^{+1.5}_{-0.7}$ & --- & 534.2/660 (80.3\%) & 0.43\\ \hline
J12489--6243 & J124846.4--623743 (4a) & $0.0^{+1.2}_{-0.0}$ &  --$0.83^{+0.76}_{-0.56}$ & $0.54^{+0.25}_{-0.11}$ & --- & 205.9/660 (28.6\%) & 1.28\\
 & J124905.3--624242 (4b) & $4.2^{+5.3}_{-2.9}$ &    $3.5^{+2.7}_{-1.9}$    & $1.3^{+400}_{-1.2}$ & --- & 103.1/660 (65.1\%) & 1.37\\ \hline
J13402--6428 & J133935.8--642537 (5a) & $1.2^{+0.9}_{-0.8}$ &    $1.0\pm 0.6$ & $0.71^{+0.16}_{-0.14}$ & --- & 314.0/660 (29.5\%) & 1.02\\
 & J133959.2--642444 (5b) & $1.7^{+2.4}_{-1.4}$ &    $1.9^{+1.5}_{-1.1}$ & $0.27^{+1.39}_{-0.12}$ & --- & 171.9/660 (60.9\%) & 1.03\\ \hline
J15293--5609 & J152929.3--561213 (6) & $0.34^{+0.28}_{-0.23}$ & $2.4^{+0.6}_{-0.5}$ & $0.47^{+0.34}_{-0.13}$ & --- & 249.8/660 (56.4\%) & 1.56\\ \hline
J15368--5102 & J153658.6--510221 (7) & $0.84^{+0.82}_{-0.61}$ & $1.3\pm 0.6$ & $0.44^{+0.14}_{-0.10}$ & --- & 269.6/660 (41.7\%) & 0.40\\ \hline
J15391--5307 & J153916.7--530815 (8) & $37^{+29}_{-19}$      & $3.2^{+3.0}_{-2.1}$    & $52^{+150,000}_{-50}$ & --- & 194.7/660 (49.4\%) & 0.96\\ \hline
J15415--5029 & J154126.4--502823 (9a) & $0.0^{+1.1}_{-0.0}$ &  --$0.43^{+0.68}_{-0.44}$ & $0.56^{+0.21}_{-0.17}$ & --- & 235.7/660 (55.5\%) & 0.44\\
 & J154140.5--502848 (9b) & $0.5^{+1.6}_{-0.5}$ &    $1.3^{+1.5}_{-1.0}$ & $0.07^{+0.13}_{-0.03}$ & --- & 101.3/660 (49.5\%) & 0.45\\ \hline
J16173--5023 & J161728.2--502242 (10a)\footnote{With a Gaussian emission line added with $E_{\rm line} = 6.8^{+0.5}_{-0.4}$~keV, $\sigma_{\rm line} = 0.45^{+0.58}_{-0.45}$~keV, and $N_{\rm line} = (9.1^{+37}_{-6.4})\times 10^{-5}$ ph~cm$^{-2}$~s$^{-1}$.} & $0.57^{+0.31}_{-0.25}$ & $0.97^{+0.34}_{-0.30}$ & $6.1^{+2.8}_{-1.5}$ & $0.34^{+0.38}_{-0.23}$ & 32.7/27 & 1.94\\ 
 & J161720.6--502415 (10b) & $22^{+39}_{-19}$    &    $1.4^{+4.1}_{-2.7}$ & $1.5^{+33}_{-0.9}$  & --- & 102.8/660 (46.1\%) & 1.95\\ \hline
J16206--5253 & J161955.4--525230 (11) & $3.8^{+3.3}_{-2.6}$ &  $0.67^{+0.98}_{-0.90}$ & $0.85^{+0.35}_{-0.19}$ & --- & 287.7/660 (55.3\%) & 0.99\\ \hline
J16413--4046 & J164119.4--404737 (12) & $6.3^{+4.1}_{-3.4}$ &    $0.47^{+0.76}_{-0.73}$ & $2.6^{+0.7}_{-0.4}$ & --- & 416.9/660 (81.4\%) & 0.41\\
 & J164119.4--404737 (12)\footnote{With a Gaussian emission line added with $E_{\rm line} = 5.0\pm 0.3$~keV, 
$\sigma_{\rm line} = 0.3\pm 0.3$~keV, and $N_{\rm line} = (1.9^{+1.7}_{-1.5})\times 10^{-5}$ ph~cm$^{-2}$~s$^{-1}$.} 
& $3.9^{+3.9}_{-2.6}$ & $0.29^{+0.94}_{-0.80}$ & $1.89\pm 0.47$ & --- & 8.5/9 & 0.41\\ \hline
J16560--4958 & J165551.9--495732 (13) & $2.3^{+0.7}_{-0.6}$ &  $2.2^{+0.5}_{-0.4}$ & $18^{+16}_{-7}$  & $0.88^{+0.12}_{-0.18}$ & 55.9/49 & 0.35\\ \hline
J21565+5948 & J215604.2+595604 (14) & $1.1^{+0.8}_{-0.6}$ &  $1.6\pm 0.5$ & $1.4^{+0.6}_{-0.3}$ & --- & 421.3/660 (77.4\%) & 0.65\\ \hline
\end{tabular}
\end{center}
\end{minipage}
\end{table*}

The two candidates with the highest count rates, CXOU~J161728.2--502242 and 
CXOU J165551.9--495732, have spectra that show evidence for photon pile-up.
When their spectra, binned with at least 20 counts per bin, are fitted with 
absorbed power-law models and no pile-up correction, the $\chi^{2}_{\nu}$ 
values obtained are both 1.8 (31 and 50 dof, respectively).  For the former 
source, a pile-up correction improves the fit to $\chi^{2}_{\nu} = 1.32$ 
(30 dof), but residuals are still seen between 6 and 7\,keV.  We added a
Gaussian emission line with an energy of $6.8^{+0.5}_{-0.4}$\,keV, and the
fit improves further to $\chi^{2}_{\nu} = 1.21$ (27 dof).  While the 
statistical significance of this line is also marginal, we report the fit
parameters with the line in Table~\ref{tab:spectra}.  For
CXOU~J165551.9--495732, the fit improves to $\chi^{2}_{\nu} = 1.1$ (49 dof)
with the pile-up correction, and the residuals do not suggest that other
spectral components are required.

\subsection{Assessing the Likelihood that the Candidate Counterparts are
the True Counterparts of the IGR Sources}

With the spectral fits completed, we used Equation~1 to estimate the 
probabilities of spurious associations between the {\em Chandra} sources 
listed in Tables~\ref{tab:counterparts} and \ref{tab:spectra} and their 
respective IGR sources.  We used the spectral parameters given in 
Table~\ref{tab:spectra} to determine the absorbed 2--10\,keV flux for
each source.  These fluxes are given in Table~\ref{tab:spurious} along
with $N(>F_{2-10~\rm keV})$ \citep[the density of sources on the sky from][]{sugizaki01},
$\theta$, $\theta_{INTEGRAL}$, and an estimate of the probability that each 
association is spurious.  Taken as a group of 18 candidates, we would expect 
probabilities of 5.6\%, 11.1\%, and 22.2\% to correspond to one, two, and 
four (out of 18) spurious associations.  The calculated probabilities range 
from 0.35\% to 24\%, and we would, roughly, consider the four with values 
between 0.35\% and 1.7\% to be very likely associations, the three with 
values between 2.2\% and 3.0\% to be likely associations, the seven with 
values between 5.1\% and 8.8\% to be somewhat likely associations, and the 
four with values of 12\% or greater to be questionable.  For the cases 
with two candidate counterparts (i.e., cases for which we know one of the 
counterparts is spurious), the candidates with the higher probabilities of 
being spurious have values of 2.2\%, 12\%, 18\%, and 24\%.  Although it is 
somewhat surprising that one of the spurious sources has a probability of 
only 2.2\%, the other three would fall in the ``questionable'' category.

The faintest sources actually become significantly more likely to be 
spurious when one considers that the AGN contribution causes the 
$\log{N}$-$\log{S}$ to steepen at low flux levels.  Although it is 
clear that this steepening occurs \citep{hands04,kim07}, there are
complications in the Galactic plane due to absorption of the background
AGN by the ISM as well as uncertainties about the decomposition of the 
$\log{N}$-$\log{S}$ into its Galactic and extragalactic parts.  However, 
if we take the combination of Galactic and extragalactic components 
estimated by \cite{sugizaki01}, the spurious probabilities for the 
four faintest sources increase by factors of 1.7--2.2.  The changes 
are much smaller for the brighter sources, but, in Table~\ref{tab:spurious},
we provide the spurious probabilities with the extragalactic component 
included for all of the {\em Chandra} candidates.

\begin{table*}
\caption{Quantities for Calculating the Probability of Spurious Associations\label{tab:spurious}}
\begin{minipage}{\linewidth}
\begin{center}
\begin{tabular}{llcccccc} \hline \hline
  &  &  & $N(>F_{2-10~\rm keV})$ & $\theta$ & $\theta_{INTEGRAL}$ & Probability & Probability\\
IGR Name & CXOU Name & $F_{2-10~\rm keV}$\footnote{The absorbed 2--10 keV flux in units of erg~cm$^{-2}$~s$^{-1}$.} & (degrees$^{-2}$)\footnote{The X-ray source density at the flux level for each source from the {\em ASCA} survey of the Galactic plane using the $\log{N}$-$\log{S}$ from \S\,3.1 \citep{sugizaki01}.} & (arcmin)\footnote{The angular distance between the center of the {\em INTEGRAL} error circle and the source.} & (arcmin)\footnote{The 90\% confidence {\em INTEGRAL} error radius \citep{bird10}.} &  (\%)\footnote{Probability of a spurious association using the $\log{N}$-$\log{S}$ from \S\,3.1.} & (\%)\footnote{Probability of a spurious association using a $\log{N}$-$\log{S}$ that is the sum of the relation given in \S\,3.1 and a steeper relation that is an estimate for the extragalactic contribution seen through the Galaxy (see \S\,3.3).}\\ \hline \hline
J01545+6437 & J015435.2+643757 (1)  & $2.94\times 10^{-13}$ & 3.83 & 1.32 & 4.1 &  5.6 & 8.0\\ \hline
J03103+5706 & J031010.6+570712 (2)  & $2.54\times 10^{-13}$ & 4.30 & 1.37 & 4.9 &  8.8 & 12.8\\ \hline
J09189--4418 & J091858.8--441830 (3) & $2.30\times 10^{-12}$ & 0.77 & 0.64 & 4.5 &  1.4 & 1.5\\ \hline
J12489--6243 & J124846.4--623743 (4a) & $5.29\times 10^{-13}$ & 2.37 & 5.42 & 4.1 &  6.1 & 7.9\\
 & J124905.3--624242 (4b) & $4.14\times 10^{-14}$ & 13.4 & 1.40 & 4.1 & 24 & 53\\ \hline
J13402--6428 & J133935.8--642537 (5a) & $5.59\times 10^{-13}$ & 2.27 & 5.02 & 3.5 &  5.1 & 6.4\\
 & J133959.2--642444 (5b) & $1.18\times 10^{-13}$ & 6.82 & 4.29 & 3.5 & 12 & 21\\ \hline
J15293--5609 & J152929.3--561213 (6) & $1.30\times 10^{-13}$ & 6.79 & 3.08 & 5.4 & 17 & 29\\ \hline
J15368--5102 & J153658.6--510221 (7) & $3.22\times 10^{-13}$ & 3.51 & 1.82 & 4.9 &  7.4 & 10.3\\ \hline
J15391--5307 & J153916.7--530815 (8) & $8.59\times 10^{-13}$ & 1.67 & 2.25 & 4.5 &  2.9 & 3.5\\ \hline
J15415--5029 & J154126.4--502823 (9a) & $5.46\times 10^{-13}$ & 2.38 & 1.12 & 3.8 &  3.0 & 3.8\\
 & J154140.5--502848 (9b) & $5.11\times 10^{-14}$ & 14.1 & 2.02 & 3.8 & 18 & 39\\ \hline
J16173--5023 & J161728.2--502242 (10a) & $5.87\times 10^{-12}$ & 0.36 & 2.11 & 3.3 &  0.35 & 0.37\\ 
 & J161720.6--502415 (10b) & $5.83\times 10^{-13}$ & 2.15 & 1.38 & 3.3 &  2.2 & 2.7\\ \hline
J16206--5253 & J161955.4--525230 (11) & $6.75\times 10^{-13}$ & 1.68 & 6.09 & 4.6 &  6.4 & 7.9\\ \hline
J16413--4056 & J164119.4--404737 (12) & $1.74\times 10^{-12}$ & 0.85 & 0.81 & 4.5 &  1.7 & 1.9\\ \hline
J16560--4958 & J165551.9--495732 (13) & $5.65\times 10^{-12}$ & 0.32 & 0.99 & 4.1 &  0.56 & 0.59\\ \hline
J21565+5948 & J215604.2+595604 (14) & $8.63\times 10^{-13}$ & 1.46 & 7.88 & 4.9 &  8.7 & 10.4\\ \hline
\end{tabular}
\end{center}
\end{minipage}
\end{table*}

We also considered whether extrapolations of the {\em Chandra} spectra are 
consistent with the 20--40\,keV {\em INTEGRAL} fluxes.  To estimate the 
maximum extrapolated hard X-ray flux, we fixed the photon index to the
lowest (i.e., hardest) value within its 90\% confidence error range, 
refitted the spectra, and then calculated the 20--40\,keV flux.  The
same procedure was used to determine the minimum flux, but the photon
index was fixed to the highest (i.e., softest) value.  Table~\ref{tab:integral} 
includes the 20--40\,keV fluxes reported from {\em INTEGRAL} observations 
\citep{bird10} along with an indication of whether each source has variable
hard X-ray emission, the minimum and maximum extrapolated fluxes (with the 
values of the photon index used), and a conclusion about whether the measured
{\em INTEGRAL} and extrapolated {\em Chandra} fluxes are consistent with
one another.  The fluxes are labeled either as ``consistent'' or 
``cutoff needed'' in 13 cases, where the latter conclusion indicates that
the {\em Chandra} extrapolation is higher than the {\em INTEGRAL} flux.
Three cases are labeled as ``inconsistent'' because the {\em Chandra}
extrapolation is lower than the {\em INTEGRAL} flux.  Two of these 
(CXOU~J124905.3--624242 and CXOU~J154140.5--502848) are cases with two 
candidate counterparts and high (18\% and 24\%) spurious probabilities.
The third is CXOU~J153658.6--510221 for which the {\em INTEGRAL} flux is 
$(3.8\pm 0.8)\times 10^{-12}$\,erg\,cm$^{-2}$\,s$^{-1}$ while the maximum
{\em Chandra} extrapolated flux is $1.8\times 10^{-12}$\,erg\,cm$^{-2}$\,s$^{-1}$.  
Given that this is only a factor of $\sim$2 low and that the probability for 
the association to be spurious is 7.4\% (see Table~\ref{tab:spurious}), this
should still be considered as a candidate counterpart.  

In the final two cases (CXOU~J031010.6+570712 and CXOU~J152929.3--561213), 
no conclusion is reached because an upper limit on the 20--40\,keV flux 
is quoted in \cite{bird10}.  This occurs because the {\em INTEGRAL} detection 
occurred in a different energy band.  For CXOU~J031010.6+570712, the 
{\em INTEGRAL} flux limit is $<$$3.8\times 10^{-12}$\,erg\,cm$^{-2}$\,s$^{-1}$ 
while the maximum {\em Chandra} extrapolated flux is 
$2.9\times 10^{-12}$\,erg\,cm$^{-2}$\,s$^{-1}$.  Thus, the fluxes are formally
consistent, and the association remains plausible.  For CXOU~J152929.3--561213, 
although \cite{bird10} report an upper limit, they also indicate that the source
is variable and quote a maximum 20--40\,keV flux of 
$2.3\times 10^{-12}$ \,erg\,cm$^{-2}$\,s$^{-1}$.  The maximum {\em Chandra}
extrapolated flux is 25 times lower than this, but since the source is known
to be variable, we cannot rule out the association.

In summary, based on the probabilities and the flux comparisons to {\em INTEGRAL},
it is likely that the CXOU~J124905.3--624242 and CXOU~J154140.5--502848 associations 
are spurious.  Also, due to the presence of multiple candidate counterparts, 
we know that one of the IGR~J13402--6428 and one of the IGR~J16173--5023 candidate
counterparts are spurious (unless {\em INTEGRAL} is actually detecting unresolved
flux from both of the {\em Chandra} sources in these cases).  

\begin{table*}
\caption{Comparison to {\em INTEGRAL} Fluxes\label{tab:integral}}
\begin{minipage}{\linewidth}
\begin{center}
\begin{tabular}{lclccc} \hline \hline
IGR Name & $\log{F_{INTEGRAL}}$\footnote{The base-10 logarithm of the 20--40\,keV {\em INTEGRAL} flux reported for the source in \cite{bird10}.  Sources with evidence for 20--40\,keV variability are indicated with a ``V'' in parentheses.} & CXOU Name & $\log{F_{\rm ext,min}}$ ($\Gamma$)\footnote{Minimum 20--40\,keV flux extrapolated from the {\em Chandra} spectrum using the value of the photon index ($\Gamma$) indicated.} & $\log{F_{\rm ext,max}}$ ($\Gamma$)\footnote{Maximum 20--40\,keV flux extrapolated from the {\em Chandra} spectrum using the value of the photon index ($\Gamma$) indicated.} & Conclusion\\ \hline \hline
J01545+6437 & --11.34 & J015435.2+643757 (1) & --11.71 (0.16)  & --10.64 (--0.99) & Consistent\\ \hline
J03103+5706 & $<$--11.42 & J031010.6+570712 (2) & --16.77 (7.3)   & --11.54 (0.41) & ---\\ \hline
J09189--4418 & --11.64 & J091858.8--441830 (3) & --11.81 (1.87)  & --11.16 (1.04) & Consistent\\ \hline
J12489--6243 & --11.42 (V) & J124846.4--623743 (4a) & --11.28 (--0.07) & --10.20 (--1.38) & Cutoff needed\\
 & & J124905.3--624242 (4b) & --17.14 (6.2)   & --13.17 (1.59) & Inconsistent\\ \hline
J13402--6428 & --11.64 (V) & J133935.8--642537 (5a) & --12.34 (1.58)  & --11.36 (0.45)   & Consistent\\
 & & J133959.2--642444 (5b) & --13.11 (1.53)  & --10.41 (--1.13) & Consistent\\ \hline
J15293--5609 & $<$--11.82 (V) & J152929.3--561213 (6) & --14.18 (2.97)  & --13.04 (1.92)   & ---\\ \hline
J15368--5102 & --11.42 & J153658.6--510221 (7) & --12.85 (1.90)  & --11.74 (0.68)   & Inconsistent\\ \hline 
J15391--5307 & --11.42 (V) & J153916.7--530815 (8) & --14.56 (6.2)   & --11.37 (1.11)   & Consistent\\ \hline
J15415--5029 & --11.34 & J154126.4--502823 (9a) & --11.28 (0.25)  & --10.52 (--0.87) & Cutoff needed\\
 & & J154140.5--502848 (9b) & --14.60 (2.83)  & --12.15 (0.26)   & Inconsistent\\ \hline
J16173--5023 & --11.28 & J161728.2--502242 (10a) & --11.25 (1.31)  & --10.55 (0.67)   & Cutoff needed\\
 & & J161720.6--502415 (10b) & --14.39 (5.4)   & --9.94 (--1.38)  & Consistent\\ \hline
J16206--5253 & $<$--11.82 & J161955.4--525230 (11) & --12.26 (1.66)  & --10.78 (--0.23) & Consistent\\ \hline
J16413--4056 & --11.52 & J164119.4--404737 (12) & --11.59 (1.23)  & --10.26 (--0.51) & Consistent\\ \hline
J16560--4958 & --11.34 & J165551.9--495732 (13) & --12.18 (2.69)  & --11.33 (1.79)   & Consistent\\ \hline
J21565+5948 & --11.64 & J215604.2+595604 (14) & --12.54 (2.11)  & --11.64 (1.09)   & Consistent\\ \hline
\end{tabular}
\end{center}
\end{minipage}
\end{table*}

\begin{figure*}
\begin{center}
\includegraphics[scale=0.9]{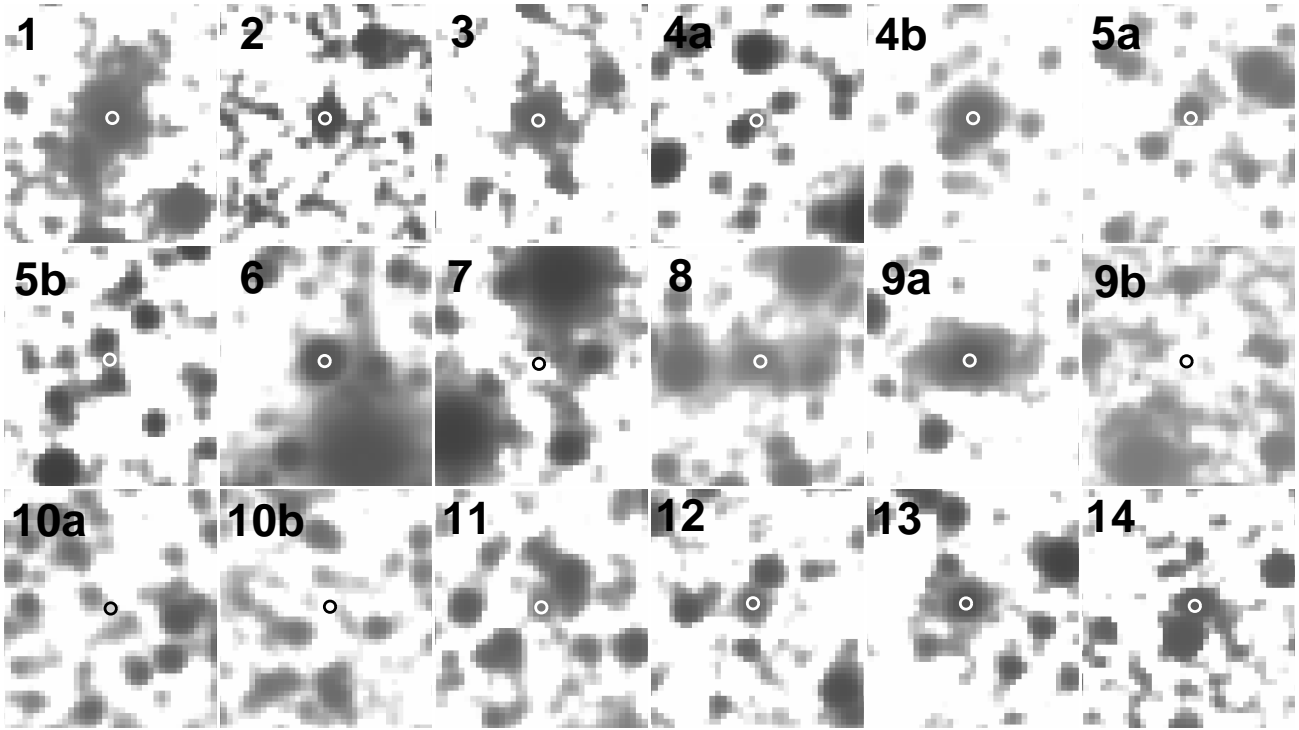}
\caption{2MASS $K_{s}$-band images for the 18 {\em Chandra} sources that are candidates for being
counterparts to the IGR sources.  For each image, North is up, and East is to the left.  The 
images are approximately 40$^{\prime\prime}$-by-40$^{\prime\prime}$.  The white and black circles mark 
the {\em Chandra} positions, and they are $1^{\prime\prime}$ in radius.  The labels correspond to 
the sources as follows: 
1=IGR~J01545+6437, 2=IGR~J03103+5706, 3=IGR~J09189--4418, 4a=IGR~J12489--6243 (candidate a), 
4b=IGR~J12489--6243 (candidate b), 5a=IGR~J13402--6428 (candidate a), 5b=IGR~J13402--6428 
(candidate b), 6=IGR~J15293--5609, 7=IGR~J15368--5102, 8=IGR~J15391--5307, 
9a=IGR~J15415--5029 (candidate a), 9b=IGR~J15415--5029 (candidate b), 
10a=IGR~J16173--5023 (candidate a), 10b=IGR~J16173--5023 (candidate b), 11=IGR~J16206--5253,
12=IGR~J16413--4046, 13=IGR~J16560--4958, and 14=IGR~J21565+5948.
\label{fig:2mass_images}}
\end{center}
\end{figure*}

\subsection{Multi-Wavelength Identifications}

We began the search for multi-wavelength counterparts by using the SIMBAD 
database\footnote{See http://simbad.u-strasbg.fr/simbad.} to determine if any 
known source coincides with each {\em Chandra} source.  We used the {\em Chandra} 
positions and a search radius that was large enough to include the position of the 
IGR source (several arcminutes in most cases as given in Table~\ref{tab:counterparts}).  
We then searched for multi-wavelength counterparts to the 18 {\em Chandra} sources in 
catalogs using the VizieR database\footnote{See http://webviz.u-strasbg.fr/viz-bin/VizieR.}.  
This search included all of the $\sim$10,000 available multi-wavelength catalogs, 
and we used the {\em Chandra} positions and a search radius of $2^{\prime\prime}$.   
This led to many identifications at optical, IR, and, in a couple cases, radio 
wavelengths.  The search included the Wide-field Infrared Survey 
Explorer\footnote{See http://irsa.ipac.caltech.edu/Missions/wise.html.} (WISE)
catalog with measurements at 3.4, 4.8, 10.2, and 23.68\,$\mu$m ($L$, $M$, $N$, 
and $O$-bands, respectively). 

The 2MASS $K_{s}$-band images are shown for the 18 {\em Chandra} sources in 
Figure~\ref{fig:2mass_images} with the {\em Chandra} positions marked.  
In 14 cases, the {\em Chandra} positions lie on near-IR sources and are
marked with white circles.  Eleven of these near-IR sources are in the 
2MASS catalog.  For the other three (CXOU~J124846.4--623743, CXOU~J133935.8--642537, 
and CXOU~J133959.2--642444, which are labeled as 4a, 5a, and 5b in the 
Figures and Tables, the {\em Chandra} positions fall on blended or 
possibly blended near-IR sources.  In four cases (CXOU~J153658.6--510221, 
CXOU~J154140.5--502848, CXOU~J161728.2--502242, and CXOU~J161720.6--502415, 
which are labeled as 7, 9b, 10a, and 10b in the Figures and Tables), the 
{\em Chandra} positions, which are marked with black circles, do not fall in 
locations with any apparent near-IR emission.

\begin{figure*}
\begin{center}
\includegraphics[scale=0.9]{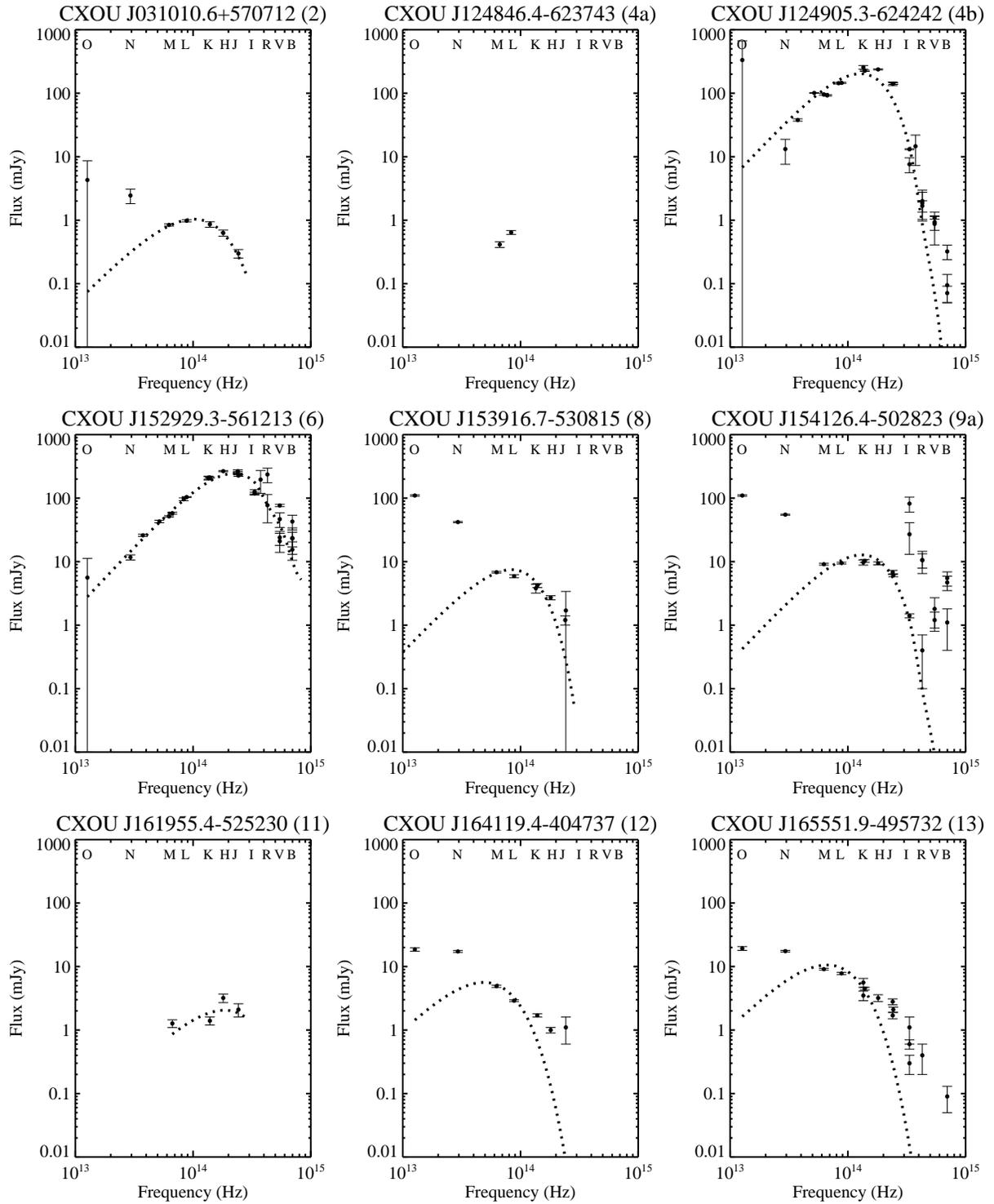}
\caption{Optical/IR spectral energy distributions (SEDs) for the sources in 
Table~\ref{tab:multiwavelength}.  The fluxes shown are not corrected for extinction.  For 
sources with more than three data points, the SEDs were fitted with a blackbody with 
extinction to determine whether they are consistent with a blackbody shape.\label{fig:sed9}}
\end{center}
\end{figure*}

In some cases, our multi-wavelength search confirms previously suggested
counterparts, and these are discussed in \S\,4.1.  We obtained new 
multi-wavelength counterparts with the {\em Chandra} positions and the
VizieR search described above for nine sources, and they are listed in 
Table~\ref{tab:multiwavelength} and discussed in Section~4.2.  While the 
table includes the optical/IR magnitudes, we converted these to 
fluxes\footnote{For cases where fluxes are not given in the catalog, we
used the magnitude-to-flux converter available at
http://www.gemini.edu/sciops/instruments/midir-resources/.}, and the spectral 
energy distributions (SEDs) are shown in Figure~\ref{fig:sed9}.  We fitted 
each SED with a blackbody with extinction to provide a rough characterization 
of its shape.  Although none of the $\chi^{2}$ values for the fits are formally 
acceptable, the fits show whether the SEDs are similar to a blackbody, as 
might be expected for a star, or if they have long-wavelength excesses, 
as might be expected for an AGN \citep{elvis94}.  The spectral 
model accounts for interstellar extinction using the analytical 
approximation of \cite{ccm89}.  The three free parameters in the fits 
are the blackbody temperature, the ratio of the blackbody radius to the 
source distance, and the number of magnitudes of extinction in the $V$-band 
($A_{V}$).

\begin{table*}
\caption{Optical/Infrared/Radio Identifications\label{tab:multiwavelength}}
\begin{minipage}{\linewidth}
\begin{center}
\begin{tabular}{cccccc} \hline \hline
Catalog\footnote{The catalogs are 
the 2 Micron All-Sky Survey \citep[2MASS and 2MASX,][]{cutri03},
the Deep Near Infrared Survey of the Southern Sky (DENIS),
the United States Naval Observatory \citep[USNO-B1.0 and USNO-A2.0,][]{monet98,monet03},
the Naval Observatory Merged Astrometric Dataset \citep[NOMAD,][]{zacharias05},
the Guide Star Catalog \citep[GSC 2.3.2,][]{lasker08},
the Third U.S. Naval Observatory CCD Astrograph Catalog \citep[UCAC3,][]{zacharias09},
the Galactic Legacy Infrared Mid-Plane Survey Extraordinaire (GLIMPSE),
the Wide-field Infrared Survey Explorer catalog \citep[WISE,][]{cutri11},
the HST Guide Star Catalog \citep[HST GSC,][]{lasker96},
the Sydney Observatory Galactic Survey \citep[SOGS,][]{fresneau07},
and the Second Epoch Molonglo Galactic Plane Survey \citep[MGPS-2,][]{murphy07}.}
& Separation\footnote{The angular separation between the {\em Chandra} position and the catalog position.} & & Magnitudes & \\ \hline \hline
\multicolumn{6}{c}{IGR J03103+5706/CXOU J031010.6+570712 (2)}\\ \hline
2MASS & $0.\!^{\prime\prime}47$ & $J = 16.82\pm 0.15$ & $H = 15.53\pm 0.12$ & $K_{s} = 14.72\pm 0.10$ & \\
WISE & $0.\!^{\prime\prime}44$ & $L = 13.670\pm 0.041$ & $M = 13.261\pm 0.042$ & $10.417\pm 0.246$ & $O > 8.348$\\ \hline
\multicolumn{6}{c}{IGR J12489--6243/CXOU J124846.4--623743 (4a)}\\ \hline
GLIMPSE & $0.\!^{\prime\prime}40$ & $m_{3.6} = 14.130\pm 0.073$ & $m_{4.5} = 14.031\pm 0.107$ & & \\ \hline
\multicolumn{6}{c}{IGR J12489--6243/CXOU J124905.3--624242 (4b)}\\ \hline
USNO-A2.0 & $0.\!^{\prime\prime}76$ & $B = 19.1\pm 0.5$ & $R = 15.5\pm 0.5$ & & \\
USNO-B1.0 & $0.\!^{\prime\prime}35$ & $R_{1} = 15.21\pm 0.3$ & $B_{2} = 19.41\pm 0.3$ & $R_{2} = 16.01\pm 0.3$ & $I = 13.85\pm 0.3$\\ 
NOMAD & $0.\!^{\prime\prime}35$ & $B = 17.77\pm 0.3$ & $V = 16.52\pm 0.3$ & & \\
GSC 2.3.2 & $0.\!^{\prime\prime}08$ & $R = 15.42\pm 0.43$ & $V = 16.59\pm 0.46$ & $I = 13.14\pm 0.44$ & \\
UCAC3 & $0.\!^{\prime\prime}12$ & $V = 16.351\pm 0.438$ & & & \\
2MASS & $0.\!^{\prime\prime}08$ & $J = 10.122\pm 0.026$ & $H = 9.090\pm 0.022$ & $K_{s} = 8.675\pm 0.024$ & \\
GLIMPSE & $0.\!^{\prime\prime}15$ & $m_{3.6} = 8.256\pm 0.035$ & $m_{4.5} = 8.161\pm 0.027$ & $m_{5.8} = 8.070\pm 0.019$ & $m_{8.0} = 8.015\pm 0.017$\\
DENIS & $0.\!^{\prime\prime}31$ & $I = 13.26\pm 0.02$ & $J = 10.14\pm 0.07$ & $K = 8.54\pm 0.07$ & \\
WISE & $0.\!^{\prime\prime}18$ & $L = 8.237\pm 0.023$ & $M = 8.120\pm 0.022$ & $N = 8.588\pm 0.382$ & $O > 3.628$\\ \hline
\multicolumn{6}{c}{IGR J15293--5609/CXOU J152929.3--561213 (6)}\\ \hline
HST GSC & $0.\!^{\prime\prime}14$ & $V = 13.00\pm 0.28$ & & & \\
USNO-B1.0 & $1.\!^{\prime\prime}32$ & $B_{2} = 12.46\pm 0.3$ & $R_{2} = 10.24\pm 0.3$ & & \\
NOMAD & $0.\!^{\prime\prime}25$ & $B = 13.13\pm 0.3$ & $V = 12.27\pm 0.3$ &  & \\
GSC 2.3.2 & $0.\!^{\prime\prime}21$ & $R = 11.45\pm 0.42$ & $B = 13.11\pm 0.40$ & $I = 10.32\pm 0.36$ & \\
UCAC3 & $0.\!^{\prime\prime}25$ & $V = 11.732\pm 0.045$ & & & \\
2MASS & $0.\!^{\prime\prime}19$ & $J = 9.620\pm 0.026$ & $H = 8.962\pm 0.022$ & $K_{s} = 8.747\pm 0.024$ & \\ 
GLIMPSE & $0.\!^{\prime\prime}20$ & $m_{3.6} = 8.650\pm 0.054$ & $m_{4.5} = 8.726\pm 0.043$ & $m_{5.8} = 8.571\pm 0.037$ & $m_{8.0} = 8.462\pm 0.031$\\
DENIS & $0.\!^{\prime\prime}25$ & $I = 10.62\pm 0.08$ & $J = 9.54\pm 0.06$ & $K = 8.71\pm 0.06$ & \\
DENIS & $0.\!^{\prime\prime}26$ & $I = 10.70\pm 0.05$ & $J = 9.59\pm 0.05$ & $K = 8.74\pm 0.07$ & \\
SOGS\footnote{Also, gives parallax as $0.64\pm 0.05$ milliarcseconds, which corresponds to a distance of $1.56\pm 0.12$ kpc.} & $0^{\prime\prime}.34$ & $V = 13.00\pm 0.3$ & $B = 13.60\pm 0.3$ & & \\
WISE & $0.\!^{\prime\prime}23$ & $L = 8.610\pm 0.028$ & $M = 8.794\pm 0.029$ & $N = 8.718\pm 0.099$ & $O > 8.059$\\ \hline
\multicolumn{6}{c}{IGR J15391--5307/CXOU J153916.7--530815 (8)}\\ \hline
2MASS & $0.\!^{\prime\prime}30$ & $J = 14.946$ & $H = 13.946\pm 0.081$ & $K_{s} = 12.990\pm 0.067$ & \\
MGPS  & $0.\!^{\prime\prime}62$ & $S_{\rm 843 MHz} = 15.1\pm 1.5$ mJy & & & \\
DENIS & $0.\!^{\prime\prime}63$ & $J = 15.41\pm 0.16$ & $K = 13.06\pm 0.15$ & & \\
WISE & $0.\!^{\prime\prime}50$ & $L = 11.819\pm 0.055$ & $M = 11.072\pm 0.039$ & $N = 7.512\pm 0.022$ & $O = 4.829\pm 0.025$\\ \hline
\multicolumn{6}{c}{IGR J15415--5029/CXOU J154126.4--502823 (9a)}\\ \hline
USNO-A2.0 & $0.\!^{\prime\prime}49$ & $B = 14.9\pm 0.5$ & $R = 13.9\pm 0.5$ & & \\
USNO-B1.0 & $0.\!^{\prime\prime}69$ & $R_{1} = 13.88\pm 0.3$ & $B_{2} = 14.73\pm 0.3$ & $R_{2} = 13.29\pm 0.3$ & $I = 11.09\pm 0.3$\\
NOMAD & $0.\!^{\prime\prime}69$ & $B = 14.73\pm 0.3$ & $V = 16.23\pm 0.3$ & & \\
GSC 2.3.2 & $0.\!^{\prime\prime}14$ & $R = 15.51\pm 0.45$ & $B = 16.46\pm 0.51$ & $V = 17.49\pm 0.62$ & $I = 12.30\pm 0.45$\\
2MASS & $0.\!^{\prime\prime}20$ & $J = 13.537\pm 0.053$ & $H = 12.594\pm 0.058$ & $K_{s} = 12.029\pm 0.051$ & \\
2MASX & $0.\!^{\prime\prime}61$ & $J_{\rm ext} = 11.947\pm 0.097$ & $H_{\rm ext} = 11.380\pm 0.100$ & $K_{s, \rm ext} = 11.009\pm 0.095$ & \\
DENIS & $0.\!^{\prime\prime}14$ & $I = 15.52\pm 0.04$ & $J = 13.51\pm 0.07$ & $K = 12.05\pm 0.10$ & \\
WISE & $0.\!^{\prime\prime}17$ & $L = 11.296\pm 0.034$ & $M = 10.755\pm 0.029$ & $N = 7.229\pm 0.020$ & $O = 4.527\pm 0.027$\\ \hline
\multicolumn{6}{c}{IGR J16206--5253/CXOU J161955.4--525230 (11)}\\ \hline
2MASS & $0.\!^{\prime\prime}21$ & $J = 14.701$ & $H = 13.749$ & $K_{s} = 14.183\pm 0.124$ & \\
GLIMPSE & $0.\!^{\prime\prime}24$ & $m_{4.5} = 12.877\pm 0.153$ & & & \\ \hline
\multicolumn{6}{c}{IGR J16413--4046/CXOU J164119.4--404737 (12)}\\ \hline
2MASS & $0.\!^{\prime\prime}19$ & $J = 15.400$ & $H = 15.081\pm 0.115$ & $K_{s} = 13.974\pm 0.066$ & \\
WISE & $0.\!^{\prime\prime}24$ & $L = 12.599\pm 0.107$ & $M = 11.436\pm 0.059$ & $N = 8.482\pm 0.040$ & $O = 6.460\pm 0.066$\\ \hline
\multicolumn{6}{c}{IGR J16560--4958/CXOU J165551.9--495732 (13)}\\ \hline
GSC 2.3.2 & $0.\!^{\prime\prime}29$ & $R = 17.04\pm 0.43$ & $B = 19.24\pm 0.41$ & $I = 15.75\pm 0.36$ & \\
2MASS & $0.\!^{\prime\prime}16$ & $J = 14.682\pm 0.087$ & $H = 13.754\pm 0.109$ & $K_{s} = 12.930\pm 0.058$ & \\
DENIS & $0.\!^{\prime\prime}40$ & $I = 16.51\pm 0.08$ & $J = 14.48\pm 0.12$ & $K = 12.64\pm 0.16$ & \\
DENIS & $0.\!^{\prime\prime}51$ & $I = 17.10\pm 0.11$ & $J = 15.04\pm 0.13$ & $K = 13.17\pm 0.17$ & \\ 
WISE & $0.\!^{\prime\prime}20$ & $L = 11.507\pm 0.035$ & $M = 10.755\pm 0.030$ & $N = 8.471\pm 0.033$ & $O = 6.423\pm 0.071$\\ \hline
\end{tabular}
\end{center}
\end{minipage}
\end{table*}

\begin{figure}
\begin{center}
\includegraphics[scale=0.45]{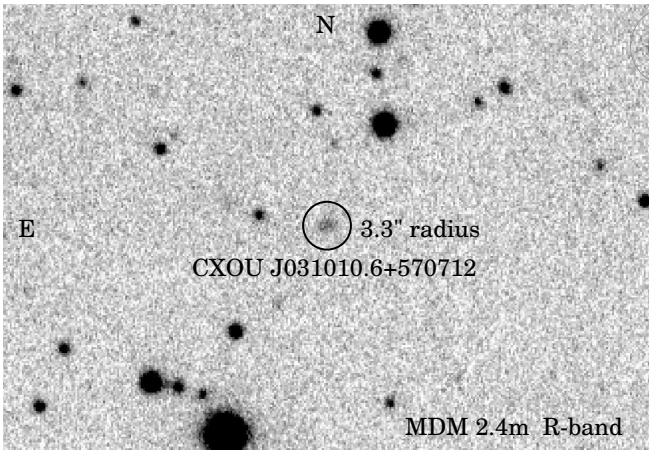}
\caption{$R$-band image in the field of IGR J03103+5706 from the MDM Observatory 2.4m Hiltner 
telescope in $1.\!^{\prime\prime}1$ seeing . The circle of radius $3.\!^{\prime\prime}3$ is centered
on the X-ray coordinates of CXOU J031010.6+570712 (2) and was used to measure the magnitude,
$R = 22.30 \pm 0.11$, for the extended (see Figure~\ref{fig:mdm_extended}) optical 
counterpart.\label{fig:mdm_image}}
\end{center}
\end{figure}

\begin{figure}
\begin{center}
\includegraphics[scale=0.45]{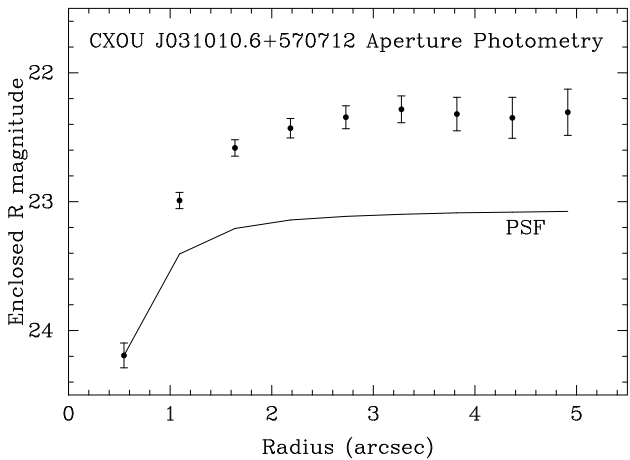}
\caption{Aperture photometry of the extended optical counterpart of CXOU J031010.6+570712 (2)
in Figure~\ref{fig:mdm_image}.  The curve-of-growth is broader than that of a PSF star.  The 
two curves are normalized at the smallest aperture used.\label{fig:mdm_extended}}
\end{center}
\end{figure}

\begin{figure*}
\begin{center}
\includegraphics[scale=0.8]{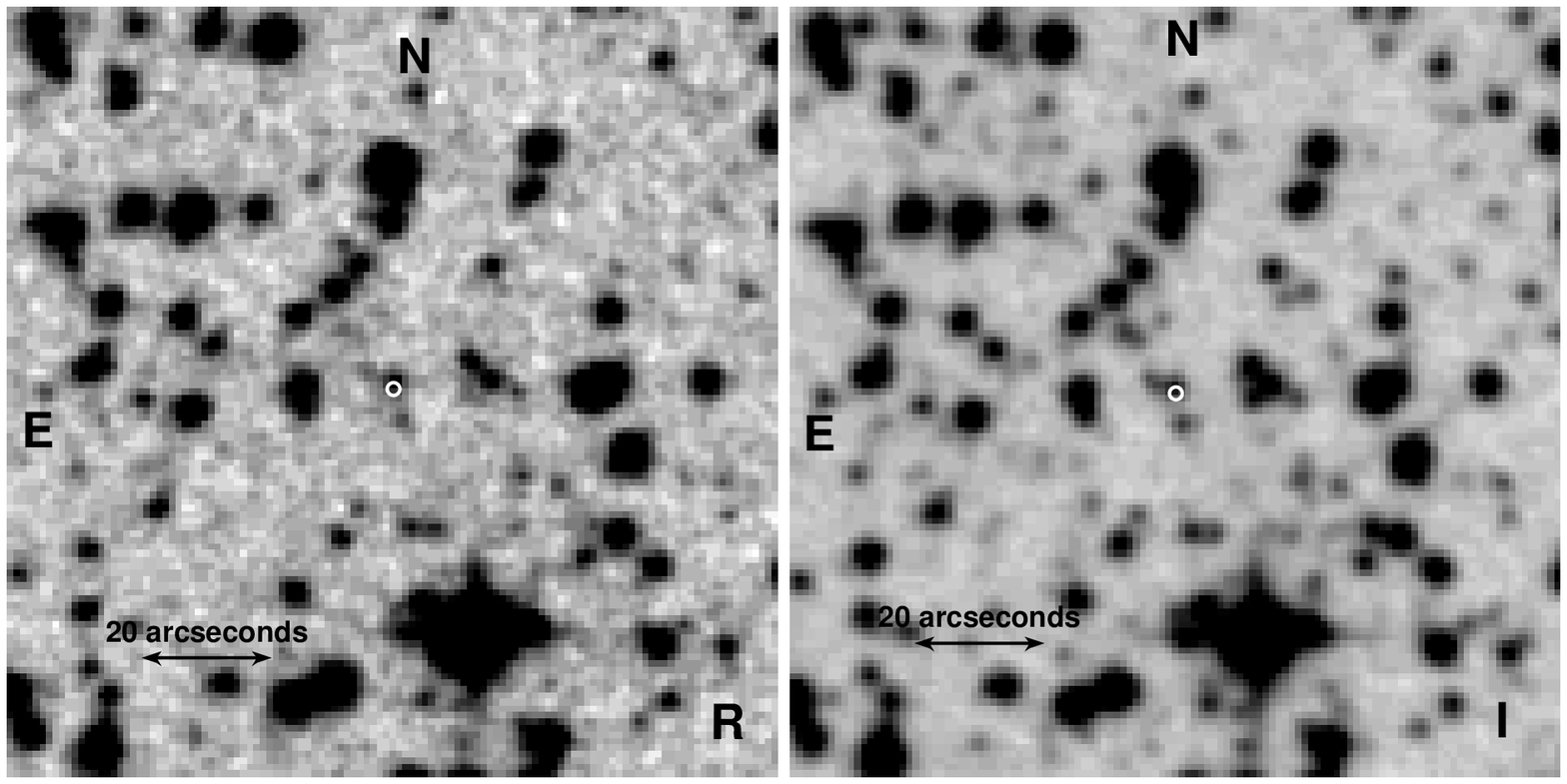}
\caption{$R$ and $I$-band images for the field around CXOU~J161728.2--502242 (10a).  The white 
circle marks the position of the {\em Chandra} source and has a radius of $1^{\prime\prime}$.  
The nearby optical source is $0.\!^{\prime\prime}67$ away in the USNO-A2.0 catalog and 
$0.\!^{\prime\prime}93$ away in the USNO-B1.0 catalog, which is margnially consistent with the 
{\em Chandra} position.  The $R$ and $I$ magnitudes are both near 17.\label{fig:dss10a}}
\end{center}
\end{figure*}

\section{Specific Results and Discussion for Each IGR Source}

\subsection{Confirmations of Previously Suggested Identifications}

IGR J01545+6437: The position of the {\em Chandra} source CXOU J015435.2+643757 (1) 
is consistent with that of LEDA~166414, which was previously suggested as the counterpart 
of the IGR source \citep{masetti8}.  Based on an optical spectrum, this source is identified 
as a Seyfert 2 AGN at a redshift of $z = 0.0355$ \citep{masetti8}.  While the {\em Chandra}
position confirms the previous identification, the spectral parameters reported in
Table~\ref{tab:spectra} ($N_{\rm H} = (0.0^{+0.4}_{-0.0})\times 10^{22}$ cm$^{-2}$, 
$\Gamma =$~--$0.40^{+0.56}_{-0.59}$) are atypical for a Seyfert 2 in that the column
density is lower and the power-law index is harder than expected.  We refitted the 
{\em Chandra} spectrum with the power-law index fixed to a value of $\Gamma = 1.8$, 
which is typical for AGN, and we obtain $N_{\rm H} = (1.8^{+1.3}_{-0.8})\times 10^{22}$ cm$^{-2}$.
While these values are consistent with what is expected for a Seyfert 2 AGN, the
C-statistic for this fit is 203.2 compared to 177.4 when $\Gamma$ was left as
a free parameter.  Using {\ttfamily goodness} to determine the percentage of simulated 
spectra with a fit statistic less than that obtained for the data gives a value of
99.5\%, indicating that the fit with $\Gamma$ fixed is poor.  We conclude that the
X-ray spectrum of this source may be unusual for a Seyfert 2 AGN, but we caution that 
the spectrum only includes $\sim$30 counts.  It would be useful to obtain a higher 
quality X-ray spectrum in the future.  For example, if a strong iron line is present, 
it could be causing the atypical value of $\Gamma$. 

IGR J09189--4418: The position of the {\em Chandra} source CXOU J091858.8--441830 (3) 
is consistent with that of 2MASX J09185877--4418302, which was previously suggested as the 
counterpart of the IGR source based on a {\em Swift}/XRT position \citep{landi10a}.  Given
that it is an extended source in the near-IR, \cite{landi10a} suggested that the source
is probably an AGN.  Our measurement of $N_{\rm H} = (4.6^{+1.1}_{-1.3})\times 10^{22}$ cm$^{-2}$
suggests that the source may be a Seyfert 2.

IGR J21565+5948:  The position of the {\em Chandra} source CXOU J215604.2+595604 (14)
is consistent with that of 2MASS J21560418+5956045, which was previously suggested as the
counterpart of the IGR source based on a {\em Swift}/XRT position \citep{landi10b}.  This 
source is also in the {\em ROSAT} catalog as 1RXS J215604.4+595607.  Follow-up optical
spectroscopy has shown that it is a Seyfert 1 AGN at a redshift of $z = 0.208$ 
\citep{bikmaev10}.

\subsection{Sources of Previously Unknown Nature with {\em Chandra} Counterparts}

\subsubsection{IGR J03103+5706}  

CXOU J031010.6+570712 (2): This source is listed in the 2MASS and WISE catalogs. In 
the SED, Figure~\ref{fig:sed9}, the near-IR points are well-described by a blackbody, 
but the 10\,$\mu$m WISE point is far from the blackbody.  We also obtained $R$-band 
images of the field of IGR~J03103+5706 on the 2.4\,m Hiltner telescope of the MDM 
Observatory on 2011 January 13 UT.  The Ohio-State Multi-Object Spectrograph
\citep{martini11} was used in imaging mode, giving a scale of 
$0.\!^{\prime\prime}273$ pixel$^{-1}$ on a University of Arizona Imaging Technology 
Laboratory $4064\times 4064$ CCD.  Five exposures of 300\,s were reduced and combined 
using standard methods; a portion of the resulting image centered on 
CXOU~J031010.6+570712 is shown in Figure~\ref{fig:mdm_image}.  The seeing was 
$1.\!^{\prime\prime}1$.  A faint optical counterpart in coincidence with the X-ray 
candidate has position R.A.=$03^{\rm h}10^{\rm m}10.\!^{\rm s}55$,
Decl.=$+57^{\circ}07^{\prime}12.\!^{\prime\prime}1$ (J2000.0) measured with respect to 
the USNO-B1.0 catalog \citep{monet03}.  The curve-of-growth from aperture photometry 
is shown in Figure~\ref{fig:mdm_extended}, which clearly shows the object to be 
extended in comparison with the stellar point-spread function (PSF). The photometry 
was calibrated using the \citet{landolt92} group of standard stars around RU~149.
We measure $R = 22.30 \pm 0.11$ in a $3.\!^{\prime\prime}3$ radius aperture, which 
includes all of the detectable light from this apparent galaxy.  The extended nature 
of the source and the 10\,$\mu$m excess likely indicate that the source is an AGN.

It should also be noted that CXOU~J031010.6+570712 is not the same source that 
was previously suggested as a possible counterpart to IGR~J03103+5706 by
\cite{maiorano11}.  The other candidate was selected for being an extended
near-IR source (2MASX J03095498+5707023) with a radio counterpart (and
therefore and AGN candidate) in the {\em INTEGRAL} error circle.  Although 
our {\ttfamily wavdetect} source detection procedure did not detect a 
{\em Chandra} source at the location of the 2MASX source, there are three 
ACIS counts within $3^{\prime\prime}$ of the 2MASX source.  Given the background 
rate estimate of 0.23\,s$^{-1}$ in a circular region with a $3^{\prime\prime}$ 
radius, 3 counts corresponds to a detection at the 99.8\% confidence (2.9-$\sigma$) 
level.  Even if this is a true detection, CXOU~J031010.6+570712 is significantly 
brighter (17.4 counts), but it still raises some uncertainty about which source
is the correct counterpart.  This is especially true because IGR~J03103+5706 was 
only detected by {\em INTEGRAL} during one satellite revolution \citep{bird10}, 
and its long-term X-ray behavior is unknown.

\subsubsection{IGR J12489--6243} 

CXOU J124846.4--623743 (4a): Although this source is not clearly resolved from a
nearby source in the 2MASS images, it becomes resolved in the GLIMPSE images.  
The positive slope of the two GLIMPSE points is consistent with a blackbody, but 
other models are certainly not ruled out. In X-rays, the source has the hardest best
fit photon index of any of our {\em Chandra} candidates, $\Gamma$ = --$0.83^{+0.76}_{-0.56}$, 
and the upper limit on the column density ($N_{\rm H} < 1.2\times 10^{22}$\,cm$^{-2}$)
is slightly less than the Galactic $N_{\rm H}$ along this line of sight.  The properties 
strongly suggest a Galactic origin, but the precise nature is unclear.  In addition, 
{\em INTEGRAL} detected hard X-ray variability from IGR~J12489--6243 (see 
Table~\ref{tab:integral}), which would also not be surprising for a Galactic source.
The most likely possibilities for the classification are that the source is a CV or 
a HMXB (an LMXB is unlikely because they typically have softer X-ray spectra).

CXOU J124905.3--624242 (4b): This is a bright optical and IR source that is
present in many of the catalogs.  The SED shown in Figure~\ref{fig:sed9} can be 
approximately described by a blackbody.  Although there are significant deviations
from a blackbody, there is no excess in the IR (e.g., below $\sim$$10^{14}$\,Hz in 
the $L$ through $O$-bands).  It is likely that the optical/IR SED is dominated by
emission from a star and that the source is Galactic.  For this source, 
Figure~\ref{fig:sed9} shows the best fit values of $A_{V}\sim 13$ and $T\sim 100,000$~K.
However, these two parameters are strongly correlated, and the temperature is 
as low as 3000~K for $A_{V} = 6$, for example.  The currently available information
does not allow us to rule out this level of extinction.  The X-ray spectral parameters
are poorly constrained, but the column density 
($N_{\rm H} = (4.2^{+5.3}_{-2.9})\times 10^{22}$\,cm$^{-2}$) either suggests that the 
source is at a relatively large distance or that it has some level of intrinsic 
absorption.  However, it is important to keep in mind that the softness of the 
power-law component makes this source less likely than candidate 4a to be the 
counterpart to the IGR source.

\subsubsection{IGR J13402--6428}

CXOU J133935.8--642537 (5a):  The {\em Chandra} position for this source falls 
$1.\!^{\prime\prime}4$ from 2MASS~J13393589-6425363, which is a large enough difference
to make the {\em Chandra}/2MASS association unlikely.  The $K_{s}$-band magnitude for
the 2MASS source is $12.53\pm 0.03$, and a better near-IR image is needed to constrain
the level of near-IR emission from the {\em Chandra} source.

CXOU J133959.2--642444 (5b):  This {\em Chandra} source falls between two 
$K_{s}\sim 14$ stars.  The {\em Chandra} position is $>$3$^{\prime\prime}$ from both
of these stars, ruling out an association with either of them, and a better near-IR
image is needed to determine if there is a fainter near-IR source between the two
stars.

\subsubsection{IGR J15293--5609}  

CXOU J152929.3--561213 (6): This is a bright optical and IR source that appears to 
be point-like.  Its SED is well-constrained from the $N$-band to the optical and
is well-described by a blackbody with no evidence for an IR excess.  The source 
appears in the Sydney Observatory Galactic Survey (SOGS) catalog \citep{fresneau07}, 
and this survey includes a parallax measurement for the source, indicating that it 
is at a distance of $1.56\pm 0.12$\,kpc.  The fact that the X-ray column density is 
$N_{\rm H} = (3.4^{+2.8}_{-2.3})\times 10^{21}$\,cm$^{-2}$, which is significantly below
the Galactic column density along this line of sight ($1.56\times 10^{22}$\,cm$^{-2}$)
is consistent with a relatively nearby source.  The 0.3--10\,keV X-ray luminosity 
is $(1.4^{+1.0}_{-0.4})\times 10^{32}$\,erg\,s$^{-1}$.

In this case, it is possible to use the blackbody fit to the SED to constrain the
spectral type of the star.  From the measured $N_{\rm H}$ value, we derive
$A_{V} = 1.5^{+1.3}_{-1.0}$ using the relationship given in \cite{go09}.  At values 
of $A_{V}$ corresponding to the minimum (0.5), the best estimate (1.5), and the
maximum (2.7), the blackbody temperatures are 4200, 5100, and 7000\,K, 
respectively.  At a distance of 1.56~kpc, the blackbody radii for these three
temperatures are 16.4, 14.5, and 12.0\,\Rsun.  The temperatures are consistent
with the star being a K or a G-type giant, but the radii are more consistent with 
an early K-type giant \citep{cox00}.  Thus, we suggest that IGR~J15293--5609 is a 
symbiotic system with a luminosity class III donor.  Given the X-ray luminosity
quoted above, the accretor is most likely to be a white dwarf.

\subsubsection{IGR J15368--5102}

CXOU J153658.6--510221 (7): No counterparts are present in any of the multi-wavelength
catalogs at the {\em Chandra} position of this source.  In addition, it is clear from
the $K_{s}$-band image (Figure~\ref{fig:2mass_images}) that any near-IR counterpart 
must be below the 2MASS detection threshold.  

\subsubsection{IGR J15391--5307}  

CXOU J153916.7--530815 (8):  We previously reported on the {\em Chandra} position and
the 2MASS association for this source \citep{tomsick11_3293}.  Based on the 2MASS
magnitudes and the very high $N_{\rm H}$ from the X-ray spectrum, we concluded that
the source probably has intrinsic absorption, which is possible either for an HMXB
or an AGN (Seyfert 2).  The new information we have obtained in this work is 
that the source is also a $15.1\pm 1.5$\,mJy radio source (see Table~\ref{tab:multiwavelength})
as well as having a strong IR component with fluxes in $N$ and $O$-band that are
orders of magnitude above the blackbody fit (see Figure~\ref{fig:sed9}).  This
strongly favors the AGN hypothesis, and this source is very likely a Seyfert 2 galaxy.

\subsubsection{IGR J15415--5029} 

CXOU J154126.4--502823 (9a): This source is in the 2MASS catalog of extended sources 
(2MASX~J15412638--5028233), and the fact that it is extended is clear from 
Figure~\ref{fig:2mass_images}.  It has also been identified with the galaxy 
LEDA~3077397/WKK98~5204 \citep{wkk01}.  The SED in Figure~\ref{fig:sed9} shows 
a strong IR excess with a rising flux all the way to the longest WISE wavelength 
($O$-band).  In addition, this source is likely associated with the $77.3\pm 2.9$\,mJy 
radio source, MGPS~J154126--502823 \citep{murphy07}.  Although the radio source is
$2.\!^{\prime\prime}2$ from the {\em Chandra} position, the rms (68\% confidence) uncertainty 
on the radio position is $1.\!^{\prime\prime}7$ and $1.\!^{\prime\prime}9$ in R.A. and Decl., 
respectively.  Thus, the {\em Chandra} and radio positions are consistent within
errors that are only slightly larger than the 1-$\sigma$ level.  This information all 
points to the source being an AGN, and it may be a Seyfert 1 since the column density 
is relatively low $N_{\rm H} < 1.1\times 10^{22}$\,cm$^{-2}$.  The large amount of scatter 
in the optical part of the SED may indicate variability.  We note that this source
fits in well with the \cite{maiorano11} efforts to identify AGN among IGR sources
using the 2MASX and radio catalogs.

CXOU J154140.5--502848 (9b): No counterparts are present in any of the multi-wavelength
catalogs at the {\em Chandra} position of this source.  In addition, it is clear from
the $K_{s}$-band image (Figure~\ref{fig:2mass_images}) that any near-IR counterpart 
must be below the 2MASS detection threshold.  Based on the X-ray information for this
source (9b) and 9a, we argue that 9a is a much more likely candidate, and the fact that 
9a is an AGN makes it even more strongly favored as being identified with IGR~J15415--5029.

\subsubsection{IGR J16173--5023}

CXOU J161728.2--502242 (10a): In this case, the catalog search was somewhat ambiguous.
The closest 2MASS source is $4.\!^{\prime\prime}2$ away from the {\em Chandra} position, 
and the sources are clearly not associated (see also Figure~\ref{fig:2mass_images}).
However, the search uncovered an optical source in the USNO catalogs that is
only $0.\!^{\prime\prime}67$ away (in USNO-A2.0) and $0.\!^{\prime\prime}93$ away
(in USNO-B1.0) from the {\em Chandra} position, and these might be associated.  
Figure~\ref{fig:dss10a} shows $R$ and $I$-band images from the Digitized Sky Survey, 
and although the images are not of sufficient quality to be certain, the association
appears to be plausible.  In USNO-B1.0, the optical magnitudes of this source are 
$B_{2} = 18.15$, $R_{1} = 17.02$, $R_{2} = 17.08$, and $I = 17.14$.  In addition, the 
SIMBAD search showed that the {\em Chandra} position is consistent with that of the 
soft X-ray (0.1--2.4\,keV) {\em ROSAT} source 1RXS~J161728.1--502238.  The photon 
index measured by {\em Chandra} for this source is relatively hard 
$\Gamma = 0.97^{+0.34}_{-0.30}$, but the {\em ROSAT} detection and the column density
of $N_{\rm H} = (5.7^{+3.1}_{-2.5})\times 10^{21}$\,cm$^{-2}$, which is significantly
less than the Galactic column density of $1.94\times 10^{22}$\,cm$^{-2}$, indicate
that the source is Galactic and that it may be relatively nearby.  The optical
counterpart, if it is truly associated with the source, as well as the possible
iron line seen in the {\em Chandra} spectrum (see Table~\ref{tab:spectra}) are
also consistent with the source being in the Galaxy.  Due to the lack of a 
near-IR counterpart, the source is not an HMXB, but its hard X-ray power-law
and its possible optical counterpart may suggest that it is a magnetic CV.

CXOU J161720.6--502415 (10b): While the closest 2MASS source to this {\em Chandra}
position is $7.\!^{\prime\prime}3$ away, the catalog search uncovered a closer source
in the GLIMPSE catalog.  This IR source is $0.\!^{\prime\prime}87$ from the {\em Chandra} 
position, which is only marginally consistent given the 90\% confidence {\em Chandra}
position uncertainty of $0.\!^{\prime\prime}64$.  Also, we note that for the {\em Chandra}
sources that do have GLIMPSE counterparts given in Table~\ref{tab:multiwavelength}, 
the separations are between $0.\!^{\prime\prime}15$ and $0.\!^{\prime\prime}40$.  While
we doubt this association, the {\em Chandra} spectrum allows for a very high column 
density, $N_{\rm H} = (2.2^{+3.9}_{-1.9})\times 10^{23}$\,cm$^{-2}$, which could suggest
that the source is an obscured HMXB or a Seyfert 2 AGN.  For both source types, an IR 
counterpart would be expected.  However, due to the uncertainty on the $N_{\rm H}$
measurement and the marginal IR counterpart association, we are not able to narrow
down the possibilities for this source.

\subsubsection{IGR J16206--5253}  

CXOU J161955.4--525230 (11): This source appears in the 2MASS and GLIMPSE catalogs.  
Although the 2MASS image (see Figure~\ref{fig:2mass_images}) shows that it is blended 
with other sources, there is no evidence that it is extended.  There are only four
points in the SED (see Figure~\ref{fig:sed9}), but it shows that the near-IR fluxes
are consistent with a blackbody (but other models are certainly not ruled out).  The 
source is not in the WISE catalog, so there is no evidence for an IR excess that could 
be the signature of an AGN.  Thus, we suspect that the source is Galactic, but the 
X-ray and multi-wavelength information do not provide any further hints to the nature 
of this source.

\subsubsection{IGR J16413--4046}  

CXOU J164119.4--404737 (12):  Previously, {\em Swift}/XRT observed the IGR~J16413--4046
field, and the presence of a soft X-ray counterpart was reported \citep{landi2731,landi3178}.
The position of the {\em Chandra} source is consistent with the {\em Swift} source, and
the refined {\em Chandra} position has provided a 2MASS identification 
\citep[see][and Table~\ref{tab:multiwavelength}]{tomsick11_3342}.  The absorbed 2--10\,keV
flux measured by {\em Swift} in 2011 January is $1.4\times 10^{-12}$\,erg\,cm$^{-2}$\,s$^{-1}$, 
and our 2011 May {\em Chandra} measurement is $(1.7\pm 0.5)\times 10^{-12}$\,erg\,cm$^{-2}$\,s$^{-1}$, 
which is consistent with no soft X-ray variability between these two observations.  The
SED (Figure~\ref{fig:sed9}) is not well-described by a blackbody.  There is a strong 
IR excess based on the WISE fluxes, and it appears that the SED has approximately 
a broken power-law shape (although it should be noted that the SED is not corrected
for extinction).  The SED suggests a non-thermal origin to the emission, which could 
indicate a jet from an AGN or an X-ray binary.  The lack of variability may suggest that 
an AGN is more likely.  Also, if the marginally-detected 5\,keV emission line is real, 
then this would make the AGN possibility certain.  For X-ray spectral fits with and
without the emission line, the column density significantly exceeds the Galactic
value of $N_{\rm H} = 4.1\times 10^{21}$\,cm$^{-2}$.  Thus, if the source is an AGN, 
it is likely a Seyfert 2.

\subsubsection{IGR J16560--4958}  

CXOU J165551.9--495732 (13): The SED for this source is very similar to 
CXOU~J164119.4--404737 with an IR excess and a broken power-law shape.  Although a
detailed modeling of the SED is beyond the scope of this paper, the apparent
curvature in the optical part of the SED (see Figure~\ref{fig:sed9}) may be due to 
extinction.  This source has the highest {\em Chandra} count rate of any source in 
our study, and we obtained a relatively good constraint on the column density, 
$N_{\rm H} = (2.3^{+0.7}_{-0.6})\times 10^{22}$\,cm$^{-2}$, which is significantly above 
the Galactic value of $N_{\rm H} = 3.5\times 10^{21}$\,cm$^{-2}$.  This source may be 
a Seyfert 2 AGN, but it was recently selected as a blazar candidate based on its IR 
colors in the WISE catalog \citep{massaro12}.  Although \cite{massaro12} did not
have a soft X-ray position for IGR~J16560--4958, we confirm that the WISE counterpart 
they identified is the same one that we identified in Table~\ref{tab:multiwavelength}.

\subsection{IGR Sources Without Likely Candidate Counterparts}

For four of the IGR sources, the {\em Chandra} observations do not result in any
candidate counterparts that stand out from the general X-ray source population.
One of these sources, IGR~J21188+4901, was found to be highly variable from the 
{\em INTEGRAL} observations \citep{bird10}.  The source was classified as one of 
the sources having variability by more than a factor of four, and \cite{bird10} 
report a peak 20--40\,keV flux of $4.5\pm 1.3$\,mCrab and an average 20--40\,keV 
flux of $<$0.2\,mCrab, indicating a difference by a factor of $>$16.  Thus, this 
source appears to be a transient.  In addition, IGR~J04069+5042 and IGR~J06552--1146 
are classified in \cite{bird10} as being variable by factors of between 1.1 and 4.  
While these relatively small factors do not prove that these sources are transients, 
the variability is likely the reason that no bright source was detected by {\em Chandra}.  
IGR~J22014+6034 is not indicated as being variable in \cite{bird10}.  If soft X-ray 
counterparts are identified in the future (e.g., during outbursts from the variable 
or transient sources), the tables of detected sources given in Appendix A will be 
useful for understanding their long-term X-ray behavior.

\begin{figure}
\begin{center}
\includegraphics[scale=0.5]{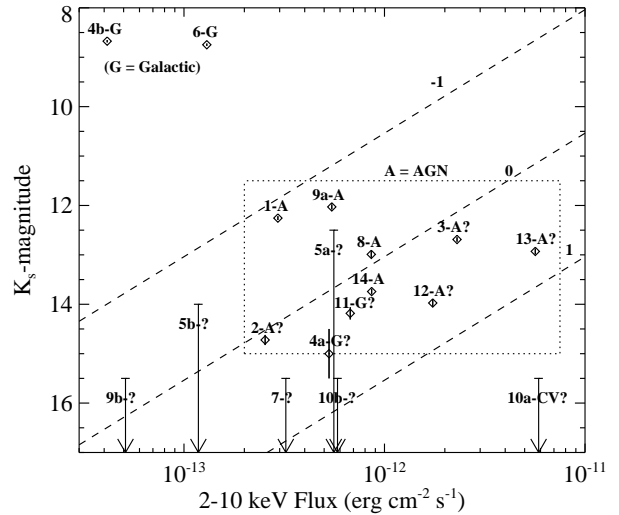}
\caption{Summary plot showing the absorbed 2--10\,keV fluxes vs. the $K_{s}$-band magnitudes for 
the {\em Chandra} sources.  The source types are A = AGN, G = Galactic, CV = Cataclysmic Variable, 
and ? = Unknown.  A question mark after one of the source types indicates some uncertainty about
that identification.  The dotted box shows the region where the eight AGN or likely AGN lie.  
The dashed lines mark constant values of $\log{(f_{X}/f_{K_{s}})}$ equal to --1, 0, and 1, and
allow for comparison to the AGN studied by \cite{watanabe04}.  The labels correspond to the 
sources as follows: 1=IGR~J01545+6437, 2=IGR~J03103+5706, 3=IGR~J09189--4418, 4a=IGR~J12489--6243 
(candidate a), 4b=IGR~J12489--6243 (candidate b), 5a=IGR~J13402--6428 (candidate a), 
5b=IGR~J13402--6428 (candidate b), 6=IGR~J15293--5609, 7=IGR~J15368--5102, 8=IGR~J15391--5307, 
9a=IGR~J15415--5029 (candidate a), 9b=IGR~J15415--5029 (candidate b), 
10a=IGR~J16173--5023 (candidate a), 10b=IGR~J16173--5023 (candidate b), 11=IGR~J16206--5253,
12=IGR~J16413--4046, 13=IGR~J16560--4958, and 14=IGR~J21565+5948.
\label{fig:summary}}
\end{center}
\end{figure}

\begin{table*}
\caption{Summary of results for the IGR Sources\label{tab:summary}}
\begin{minipage}{\linewidth}
\begin{center}
\begin{tabular}{clll} \hline \hline
              & Remaining {\em Chandra}      &               &          \\
IGR Name      & Candidate Counterparts       & Source Type   & Evidence \\ \hline\hline
J01545+6437   & CXOU J015435.2+643757 (1)    & AGN           & LEDA 166414, $z = 0.0355$\\ \hline
J03103+5706   & CXOU J031010.6+570712 (2)    & AGN?          & IR excess, Optical extension\\ \hline
J04069+5042   & --- & --- & ---\\ \hline
J06552--1146  & --- & --- & ---\\ \hline
J09189--4418  & CXOU J091858.8--441830 (3)   & AGN?          & 2MASX J09185877--4418302\\ \hline
J12489--6243  & CXOU J124846.4--623743 (4a)  & Galactic?     & SED slope, Hard X-ray spectrum\\ 
              & CXOU J124905.3--624242 (4b)  & Galactic      & Blackbody SED\\ \hline
J13402--6428  & CXOU J133935.8--642537 (5a)  & ?             & ---\\
              & CXOU J133959.2--642444 (5b)  & ?             & ---\\ \hline
J15293--5609  & CXOU J152929.3--561213 (6)   & Galactic\footnote{Based on the optical/IR spectral energy distribution and the X-ray luminosity, we suggest that this source is a symbiotic binary with an early K-type giant donor and a white dwarf accretor.}            & Parallax, Blackbody SED\\ \hline
J15368--5102  & CXOU J153658.6--510221 (7)   & ?             & ---\\ \hline
J15391--5307  & CXOU J153916.7--530815 (8)   & AGN           & Radio emission, IR excess\\ \hline
J15415--5029  & CXOU J154126.4--502823 (9a)  & AGN           & 2MASX J15412638--5028233, WKK98~5204\\ \hline
J16173--5023  & CXOU J161728.2--502242 (10a) & CV?           & Low $N_{\rm H}$, Optical counterpart?\\
              & CXOU J161720.6--502415 (10b) & ?             & ---\\ \hline
J16206--5253  & CXOU J161955.4--525230 (11)  & Galactic?     & SED slope\\ \hline
J16413--4046  & CXOU J164119.4--404737 (12)  & AGN?          & IR excess, 5\,keV Fe line?\\ \hline
J16560--4958  & CXOU J165551.9--495732 (13)  & AGN?          & IR excess\\ \hline
J21188+4901   & -- & Transient & ---\\ \hline
J21565+5948   & CXOU J215604.2+595604 (14)   & AGN           & $z = 0.208$\\ \hline
J22014+6034   & -- & -- & ---\\ \hline
\end{tabular}
\end{center}
\end{minipage}
\end{table*}

\section{Summary and Conclusions}

While we expect follow-up observations, especially optical and near-IR spectroscopy, 
to provide definite identifications of the nature of many of these sources, the
{\em Chandra} observations have allowed us to determine the nature of some of the
IGR sources as well as providing {\em Chandra} counterpart identifications for 14 
of the 18 sources.  A summary is provided in Table~\ref{tab:summary}, indicating 
that four of the IGR sources show definite or very strong evidence for being AGN.  
IGR~J01545+6437 and IGR~J21565+5948 both have redshift measurements, IGR J15391--5307 
has a strong IR excess and radio emission, and IGR~J15415--5029 is identified with 
the Galaxy WKK98~5204 (2MASX~J15412638--5028233).  IGR~J15415--5029 is one of the 
IGR sources for which we considered two {\em Chandra} candidate counterparts
(9a and 9b), but we strongly suspect that 9a is the correct counterpart.  Four other 
sources (IGR~J03103+5706, IGR~J09189--4418, IGR~J16413--4046, and IGR~J16560--4958)
are likely AGN, and we label them as ``AGN?'' in Table~\ref{tab:summary}.  For these 
sources, the evidence is extended optical or IR emission or an IR excess (or both 
features).

The remaining six IGR sources for which we obtained one or two possible {\em Chandra} 
identifications may be Galactic sources.  If the identification of IGR~J15293--5609 
with CXOU~J152929.3--561213 is correct, then this is a definite Galactic source based 
on a parallax measurement that puts it distance at $1.56\pm 0.12$~kpc.  Furthermore,
based on the temperature and radius inferred from a blackbody fit to its optical/IR
SED, it is likely a symbiotic binary with an early K-type giant donor, and its X-ray
luminosity suggests a white dwarf accretor.  IGR~J16206--5253 has {\em Chandra} and 
near-IR counterparts, and we argue from its SED that it is likely Galactic.  
IGR~J12489--6243 and IGR~J16173--5023 may be Galactic; however, there are still two 
possible {\em Chandra} counterparts for each source, and, in each case, there
is some uncertainty about the nature of at least one of the counterparts.  Once the 
nature of each of the {\em Chandra} counterparts is determined via follow-up 
measurements, it should become clear which counterpart is the most likely to be able 
to produce the hard X-ray emission detected with {\em INTEGRAL}.  IGR~J15368--5102 
has a {\em Chandra} counterpart, and IGR~J13402--6428 has two possible {\em Chandra} 
counterparts.  However, the natures of these sources are still unclear.

Figure~\ref{fig:summary} shows a plot of the $K_{s}$-magnitudes and the absorbed
2--10\,keV X-ray fluxes for the {\em Chandra} counterparts.  The eight AGN or
likely AGN fall within a range of $K_{s}$-magnitudes between 14.7 and 12.0 and within 
a range of X-ray fluxes between $2.5\times 10^{-13}$ and $5.7\times 10^{-12}$\,erg\,cm$^{-2}$\,s$^{-1}$.  
We compare these ranges to a previous study of AGN selected by 2--10\,keV fluxes measured by 
{\em ASCA}.  \cite{watanabe04} show a similar $K_{s}$ vs. 2--10\,keV X-ray flux for 
$\sim$100 {\em ASCA} detected AGN, and plot lines of constant $\log{(f_{X}/f_{K_{s}})}$.  
They find that the {\em ASCA} AGN mostly fall between values of $\log{(f_{X}/f_{K_{s}})}$
0 and 1, and Figure~\ref{fig:summary} shows that five of our AGN or likely AGN 
(2, 3, 12, 13, and 14) fall in that range.  \cite{watanabe04} also included a 
sample of {\em Chandra} AGN in their study, and a significant fraction of those
AGN have $\log{(f_{X}/f_{K_{s}})}$ values ranging from --1 to 0.  Three of our AGN 
(1, 8, 9a) fall in this range.  Thus, we conclude that the near-IR and X-ray fluxes 
that we find for our AGN are fairly typical.

Galactic sources are not expected to be well-localized in a near-IR/X-ray flux
plot, and Figure~\ref{fig:summary} shows that our Galactic or potentially
Galactic sources are spread throughout the plot.  The two sources with 
well-measured blackbody spectra are in the near-IR-bright/X-ray-faint 
corner of the diagram.  The rest of these are relatively faint in the near-IR
but have a wide range of 2--10\,keV X-ray fluxes.  We have identified IGR~J15293--5609 (6) 
as a likely symbiotic binary with an early K-type giant donor and a white dwarf accretor.
For 5a, 7, 10a, and 10b, higher angular resolution or deeper images are necessary to 
obtain optical/IR counterparts.  For 4a, 4b, 5b, and 11, optical or near-IR
spectra of the counterparts identified in this work are likely to lead to a 
determination of the source type.

\acknowledgments

JAT and AB acknowledge partial support from NASA through {\em Chandra} Award 
Number GO1-12046X issued by the {\em Chandra} X-ray Observatory Center, which 
is operated by the Smithsonian Astrophysical Observatory under NASA contract 
NAS8-03060.  We acknowledge useful comments from an anonymous referee.  This 
research has made use of the IGR Sources page maintained by J.~Rodriguez and 
A.~Bodaghee (http://irfu.cea.fr/Sap/IGR-Sources), the VizieR catalog access 
tool and the SIMBAD database, which are both operated at CDS, Strasbourg, France.  
We also acknowledge the use of the 2MASS, USNO, DENIS, and WISE catalogs.

\appendix
\begin{center}
{\bf\normalsize Appendix A}
\end{center}

IGR~J04069+5042, IGR~J06552--1146, IGR~J21188+4901, and IGR~J22014+6034 are the 
cases where the candidate counterparts did not clearly stand out from the general 
population of X-ray sources.  However, it is still possible that one of the detected
{\em Chandra} sources is the correct counterpart, and we provide the full lists of
sources detected within the ACIS-I fields-of-view using {\ttfamily wavdetect} in 
Tables~\ref{tab:j04069}, \ref{tab:j06552}, \ref{tab:j21188}, and \ref{tab:j22014}.  
The ACIS counts reported in the tables are for the full 0.3--10\,keV energy band, 
and we used the same methods described in \S~3.1 to determine the source apertures 
and the background regions.  For each {\ttfamily wavdetect} source, we used Poisson 
statistics to determine the detection significance, and we did not include sources 
with significances lower than $\sim$1.2-$\sigma$.  Using simulations, \cite{hong05}
determined an empirical formula for determining {\em Chandra} position uncertainties
that depends on the number of counts detected and the off-axis angle.  For each 
source, we include the 95\% confidence position uncertainties calculated using 
Equation 5 of \cite{hong05}.  The hardness ratios are also included in the tables
for sources with five counts or more.  We obtained optical observations of 
some of the {\em Chandra} sources in these fields, and the results will be
presented in \"Ozbey Arabac\i~et al. (2012, to be submitted).  


\clearpage

\begin{table*}
\caption{{\em Chandra} Candidate Counterparts to IGR J04069+5042\label{tab:j04069}}
\begin{minipage}{\linewidth}
\begin{center}
\begin{tabular}{ccccccc} \hline \hline
Source & $\theta$\footnote{The angular distance between the center of the {\em INTEGRAL} error circle, which is also the approximate {\em Chandra} aimpoint, and the source.} & {\em Chandra} R.A. & {\em Chandra} Decl. & ACIS  & Position &  \\
Number & (arcminutes) & (J2000)  & (J2000) &  Counts\footnote{The number of ACIS-I counts detected (after background subtraction) in the 0.3--10 keV band.} & Uncertainty\footnote{The 95\% confidence statistical uncertainty on the position using Equation 5 from \cite{hong05}.  This does not include the systematic pointing error, which is $0.\!^{\prime\prime}64$ at 90\% confidence and $1^{\prime\prime}$ at 99\% confidence.} & Hardness\footnote{The hardness is given by $(C_{2}-C_{1})/(C_{2}+C_{1})$, where $C_{2}$ is the number of counts in the 2--10 keV band and $C_{1}$ is the number of counts in the 0.3--2 keV band.}\\ \hline
1  & 0.28 & $04^{\rm h}06^{\rm m}54^{\rm s}.30$ & +$50^{\circ}41^{\prime}53.\!^{\prime\prime}0$ &  5.4 & $0.\!^{\prime\prime}53$ & --$0.07\pm 0.78$\\ 
2  & 2.80 & $04^{\rm h}06^{\rm m}48^{\rm s}.69$ & +$50^{\circ}39^{\prime}31.\!^{\prime\prime}2$ &  3.4 & $1.\!^{\prime\prime}17$ & ---\\ 
3  & 3.19 & $04^{\rm h}06^{\rm m}43^{\rm s}.99$ & +$50^{\circ}44^{\prime}46.\!^{\prime\prime}2$ &  5.4 & $1.\!^{\prime\prime}02$ & --$1.19\pm 1.10$\\ 
4  & 4.70 & $04^{\rm h}06^{\rm m}25^{\rm s}.94$ & +$50^{\circ}41^{\prime}21.\!^{\prime\prime}1$ &  5.5 & $1.\!^{\prime\prime}78$ & --$0.64\pm 0.95$\\ 
5  & 5.76 & $04^{\rm h}07^{\rm m}17^{\rm s}.11$ & +$50^{\circ}46^{\prime}42.\!^{\prime\prime}9$ &  3.5 & $3.\!^{\prime\prime}97$ & ---\\ 
6  & 5.82 & $04^{\rm h}06^{\rm m}58^{\rm s}.87$ & +$50^{\circ}47^{\prime}54.\!^{\prime\prime}8$ &  3.5 & $4.\!^{\prime\prime}07$ & ---\\ 
7  & 6.35 & $04^{\rm h}06^{\rm m}15^{\rm s}.91$ & +$50^{\circ}40^{\prime}49.\!^{\prime\prime}8$ &  3.5 & $5.\!^{\prime\prime}08$ & ---\\ 
8  & 6.45 & $04^{\rm h}07^{\rm m}07^{\rm s}.51$ & +$50^{\circ}48^{\prime}16.\!^{\prime\prime}1$ &  6.5 & $3.\!^{\prime\prime}08$ & +$0.54\pm 0.80$\\ 
9  & 6.85 & $04^{\rm h}07^{\rm m}34^{\rm s}.23$ & +$50^{\circ}45^{\prime}04.\!^{\prime\prime}5$ & 13.3 & $2.\!^{\prime\prime}03$ & +$0.73\pm 0.53$\\ 
10 & 7.63 & $04^{\rm h}06^{\rm m}45^{\rm s}.95$ & +$50^{\circ}34^{\prime}37.\!^{\prime\prime}7$ &  4.3 & $6.\!^{\prime\prime}97$ & ---\\ 
11 & 8.58 & $04^{\rm h}06^{\rm m}48^{\rm s}.17$ & +$50^{\circ}50^{\prime}37.\!^{\prime\prime}5$ & 10.3 & $4.\!^{\prime\prime}33$ & --$0.13\pm 0.55$\\ 
12 & 9.16 & $04^{\rm h}05^{\rm m}59^{\rm s}.00$ & +$50^{\circ}44^{\prime}17.\!^{\prime\prime}3$ &  3.3 & $15.\!^{\prime\prime}80$ & ---\\ 
13 & 9.54 & $04^{\rm h}07^{\rm m}38^{\rm s}.06$ & +$50^{\circ}35^{\prime}25.\!^{\prime\prime}6$ & 24.3 & $2.\!^{\prime\prime}89$ & --$1.21\pm 0.43$\\ \hline 
\end{tabular}
\end{center}
\end{minipage}
\end{table*}

\begin{table*}
\caption{{\em Chandra} Candidate Counterparts to IGR J06552--1146\label{tab:j06552}}
\begin{minipage}{\linewidth}
\begin{center}
\begin{tabular}{ccccccc} \hline \hline
Source & $\theta$\footnote{The angular distance between the center of the {\em INTEGRAL} error circle, which is also the approximate {\em Chandra} aimpoint, and the source.} & {\em Chandra} R.A. & {\em Chandra} Decl. & ACIS  & Position & \\
Number & (arcminutes) & (J2000)  & (J2000) &  Counts\footnote{The number of ACIS-I counts detected (after background subtraction) in the 0.3--10 keV band.} & Uncertainty\footnote{The 95\% confidence statistical uncertainty on the position using Equation 5 from \cite{hong05}.  This does not include the systematic pointing error, which is $0.\!^{\prime\prime}64$ at 90\% confidence and $1^{\prime\prime}$ at 99\% confidence.} & Hardness\footnote{The hardness is given by $(C_{2}-C_{1})/(C_{2}+C_{1})$, where $C_{2}$ is the number of counts in the 2--10 keV band and $C_{1}$ is the number of counts in the 0.3--2 keV band.}\\ \hline
 1 & 2.99 & $06^{\rm h}55^{\rm m}11^{\rm s}.78$ & --$11^{\circ}49^{\prime}09.\!^{\prime\prime}6$ &  2.4 & $1.\!^{\prime\prime}61$ & ---\\ 
 2 & 3.03 & $06^{\rm h}54^{\rm m}58^{\rm s}.66$ & --$11^{\circ}45^{\prime}02.\!^{\prime\prime}0$ &  3.4 & $1.\!^{\prime\prime}28$ & ---\\ 
 3 & 3.33 & $06^{\rm h}55^{\rm m}23^{\rm s}.65$ & --$11^{\circ}46^{\prime}01.\!^{\prime\prime}1$ & 13.4 & $0.\!^{\prime\prime}68$ & --$0.62\pm 0.46$\\ 
 4 & 3.34 & $06^{\rm h}55^{\rm m}20^{\rm s}.25$ & --$11^{\circ}48^{\prime}25.\!^{\prime\prime}3$ &  5.4 & $1.\!^{\prime\prime}07$ & +$0.68\pm 0.89$\\ 
 5 & 3.43 & $06^{\rm h}54^{\rm m}56^{\rm s}.31$ & --$11^{\circ}45^{\prime}34.\!^{\prime\prime}6$ &  2.4 & $1.\!^{\prime\prime}93$ & ---\\ 
 6 & 3.65 & $06^{\rm h}55^{\rm m}04^{\rm s}.35$ & --$11^{\circ}42^{\prime}49.\!^{\prime\prime}5$ &  5.4 & $1.\!^{\prime\prime}20$ & +$0.31\pm 0.80$\\ 
 7 & 3.67 & $06^{\rm h}55^{\rm m}24^{\rm s}.51$ & --$11^{\circ}45^{\prime}11.\!^{\prime\prime}9$ &  3.4 & $1.\!^{\prime\prime}66$ & ---\\ 
 8 & 3.96 & $06^{\rm h}55^{\rm m}05^{\rm s}.48$ & --$11^{\circ}42^{\prime}24.\!^{\prime\prime}2$ &  2.4 & $2.\!^{\prime\prime}44$ & ---\\ 
 9 & 4.10 & $06^{\rm h}54^{\rm m}57^{\rm s}.70$ & --$11^{\circ}43^{\prime}26.\!^{\prime\prime}2$ &  2.5 & $2.\!^{\prime\prime}52$ & ---\\ 
10 & 4.15 & $06^{\rm h}54^{\rm m}58^{\rm s}.31$ & --$11^{\circ}49^{\prime}11.\!^{\prime\prime}3$ &  8.5 & $1.\!^{\prime\prime}09$ & --$0.75\pm 0.70$\\ 
11 & 4.43 & $06^{\rm h}55^{\rm m}02^{\rm s}.70$ & --$11^{\circ}50^{\prime}14.\!^{\prime\prime}6$ &  7.5 & $1.\!^{\prime\prime}30$ & +$0.35\pm 0.67$\\ 
12 & 4.63 & $06^{\rm h}55^{\rm m}05^{\rm s}.95$ & --$11^{\circ}41^{\prime}40.\!^{\prime\prime}8$ &  8.5 & $1.\!^{\prime\prime}29$ & +$0.90\pm 0.74$\\ 
13 & 5.03 & $06^{\rm h}55^{\rm m}24^{\rm s}.96$ & --$11^{\circ}49^{\prime}40.\!^{\prime\prime}2$ &  9.5 & $1.\!^{\prime\prime}38$ & --$0.14\pm 0.54$\\ 
14 & 5.46 & $06^{\rm h}55^{\rm m}04^{\rm s}.96$ & --$11^{\circ}51^{\prime}30.\!^{\prime\prime}6$ & 10.5 & $1.\!^{\prime\prime}50$ & +$0.92\pm 0.64$\\ 
15 & 5.46 & $06^{\rm h}54^{\rm m}55^{\rm s}.96$ & --$11^{\circ}50^{\prime}25.\!^{\prime\prime}7$ &  7.5 & $1.\!^{\prime\prime}90$ & +$0.35\pm 0.67$\\ 
16 & 5.53 & $06^{\rm h}55^{\rm m}29^{\rm s}.56$ & --$11^{\circ}49^{\prime}00.\!^{\prime\prime}1$ & 35.5 & $0.\!^{\prime\prime}81$ & +$0.02\pm 0.22$\\ 
17 & 5.80 & $06^{\rm h}55^{\rm m}32^{\rm s}.52$ & --$11^{\circ}44^{\prime}20.\!^{\prime\prime}7$ &  2.5 & $5.\!^{\prime\prime}40$ & ---\\ 
18 & 5.89 & $06^{\rm h}55^{\rm m}23^{\rm s}.26$ & --$11^{\circ}41^{\prime}16.\!^{\prime\prime}4$ &  2.5 & $5.\!^{\prime\prime}61$ & ---\\ 
19 & 6.17 & $06^{\rm h}55^{\rm m}34^{\rm s}.57$ & --$11^{\circ}44^{\prime}44.\!^{\prime\prime}0$ &  4.5 & $3.\!^{\prime\prime}78$ & ---\\ 
20 & 6.22 & $06^{\rm h}55^{\rm m}32^{\rm s}.00$ & --$11^{\circ}43^{\prime}03.\!^{\prime\prime}7$ & 16.5 & $1.\!^{\prime\prime}44$ & --$0.14\pm 0.36$\\ 
21 & 6.22 & $06^{\rm h}54^{\rm m}48^{\rm s}.69$ & --$11^{\circ}49^{\prime}33.\!^{\prime\prime}7$ &  8.5 & $2.\!^{\prime\prime}28$ & --$0.75\pm 0.70$\\ 
22 & 6.28 & $06^{\rm h}54^{\rm m}49^{\rm s}.20$ & --$11^{\circ}42^{\prime}33.\!^{\prime\prime}4$ &  4.5 & $3.\!^{\prime\prime}95$ & ---\\ 
23 & 6.43 & $06^{\rm h}55^{\rm m}24^{\rm s}.04$ & --$11^{\circ}40^{\prime}45.\!^{\prime\prime}5$ &  7.5 & $2.\!^{\prime\prime}71$ & --$0.98\pm 0.84$\\ 
24 & 6.57 & $06^{\rm h}55^{\rm m}15^{\rm s}.12$ & --$11^{\circ}39^{\prime}44.\!^{\prime\prime}9$ & 15.5 & $1.\!^{\prime\prime}67$ & --$0.73\pm 0.45$\\ 
25 & 8.46 & $06^{\rm h}55^{\rm m}29^{\rm s}.25$ & --$11^{\circ}53^{\prime}14.\!^{\prime\prime}3$ &  7.4 & $5.\!^{\prime\prime}62$ & --$0.82\pm 0.88$\\ \hline 
\end{tabular}
\end{center}
\end{minipage}
\end{table*}

\begin{table*}
\caption{{\em Chandra} Candidate Counterparts to IGR J21188+4901\label{tab:j21188}}
\begin{minipage}{\linewidth}
\begin{center}
\begin{tabular}{ccccccc} \hline \hline
Source & $\theta$\footnote{The angular distance between the center of the {\em INTEGRAL} error circle, which is also the approximate {\em Chandra} aimpoint, and the source.} & {\em Chandra} R.A. & {\em Chandra} Decl. & ACIS  & Position & \\
Number & (arcminutes) & (J2000)  & (J2000) &  Counts\footnote{The number of ACIS-I counts detected (after background subtraction) in the 0.3--10 keV band.} & Uncertainty\footnote{The 95\% confidence statistical uncertainty on the position using Equation 5 from \cite{hong05}.  This does not include the systematic pointing error, which is $0.\!^{\prime\prime}64$ at 90\% confidence and $1^{\prime\prime}$ at 99\% confidence.} & Hardness\footnote{The hardness is given by $(C_{2}-C_{1})/(C_{2}+C_{1})$, where $C_{2}$ is the number of counts in the 2--10 keV band and $C_{1}$ is the number of counts in the 0.3--2 keV band.}\\ \hline
 1 & 3.26 & $21^{\rm h}18^{\rm m}39^{\rm s}.09$ & +$48^{\circ}58^{\prime}05.\!^{\prime\prime}5$ &  4.3 & $1.\!^{\prime\prime}20$ & ---\\ 
 2 & 4.31 & $21^{\rm h}18^{\rm m}47^{\rm s}.81$ & +$49^{\circ}05^{\prime}19.\!^{\prime\prime}6$ &  5.3 & $1.\!^{\prime\prime}57$ & --$0.83\pm 0.97$\\ 
 3 & 4.53 & $21^{\rm h}18^{\rm m}20^{\rm s}.23$ & +$49^{\circ}00^{\prime}43.\!^{\prime\prime}0$ &  4.1 & $2.\!^{\prime\prime}07$ & ---\\ 
 4 & 4.83 & $21^{\rm h}18^{\rm m}54^{\rm s}.59$ & +$48^{\circ}56^{\prime}19.\!^{\prime\prime}2$ &  6.1 & $1.\!^{\prime\prime}74$ & +$0.56\pm 0.85$\\ 
 5 & 5.64 & $21^{\rm h}19^{\rm m}22^{\rm s}.04$ & +$49^{\circ}00^{\prime}34.\!^{\prime\prime}2$ &  3.1 & $4.\!^{\prime\prime}18$ & ---\\ 
 6 & 5.91 & $21^{\rm h}18^{\rm m}14^{\rm s}.22$ & +$49^{\circ}03^{\prime}11.\!^{\prime\prime}8$ &  3.1 & $4.\!^{\prime\prime}70$ & ---\\ 
 7 & 6.46 & $21^{\rm h}18^{\rm m}52^{\rm s}.15$ & +$49^{\circ}07^{\prime}26.\!^{\prime\prime}3$ &  7.1 & $2.\!^{\prime\prime}87$ & +$0.62\pm 0.77$\\ 
 8 & 6.79 & $21^{\rm h}18^{\rm m}22^{\rm s}.86$ & +$49^{\circ}06^{\prime}26.\!^{\prime\prime}7$ &  7.1 & $3.\!^{\prime\prime}24$ & --$0.78\pm 0.82$\\ 
 9 & 6.87 & $21^{\rm h}18^{\rm m}39^{\rm s}.76$ & +$48^{\circ}54^{\prime}16.\!^{\prime\prime}7$ &  5.5 & $4.\!^{\prime\prime}16$ & --$1.38\pm 1.48$\\ 
10 & 7.05 & $21^{\rm h}19^{\rm m}17^{\rm s}.20$ & +$48^{\circ}55^{\prime}53.\!^{\prime\prime}1$ &  3.5 & $6.\!^{\prime\prime}75$ & ---\\ 
11 & 7.29 & $21^{\rm h}18^{\rm m}29^{\rm s}.41$ & +$48^{\circ}54^{\prime}22.\!^{\prime\prime}7$ &  2.5 & $10.\!^{\prime\prime}15$ & ---\\ 
12 & 7.95 & $21^{\rm h}18^{\rm m}13^{\rm s}.45$ & +$48^{\circ}55^{\prime}24.\!^{\prime\prime}5$ & 10.5 & $3.\!^{\prime\prime}48$ & +$0.14\pm 0.55$\\ 
13 & 8.04 & $21^{\rm h}18^{\rm m}03^{\rm s}.40$ & +$49^{\circ}04^{\prime}26.\!^{\prime\prime}9$ &  6.5 & $5.\!^{\prime\prime}48$ & +$1.46\pm 1.32$\\ 
14 & 8.62 & $21^{\rm h}19^{\rm m}40^{\rm s}.33$ & +$49^{\circ}01^{\prime}17.\!^{\prime\prime}6$ &  8.5 & $5.\!^{\prime\prime}21$ & +$0.88\pm 0.82$\\ 
15 & 8.96 & $21^{\rm h}18^{\rm m}24^{\rm s}.93$ & +$48^{\circ}52^{\prime}53.\!^{\prime\prime}2$ &  5.5 & $8.\!^{\prime\prime}85$ & +$0.82\pm 1.15$\\ 
16 & 8.97 & $21^{\rm h}18^{\rm m}58^{\rm s}.24$ & +$49^{\circ}09^{\prime}49.\!^{\prime\prime}3$ &  6.5 & $7.\!^{\prime\prime}54$ & +$0.23\pm 0.82$\\ 
17 & 9.74 & $21^{\rm h}19^{\rm m}29^{\rm s}.26$ & +$49^{\circ}07^{\prime}59.\!^{\prime\prime}9$ &  3.5 & $18.\!^{\prime\prime}20$ & ---\\ \hline 
\end{tabular}
\end{center}
\end{minipage}
\end{table*}

\begin{table*}
\caption{{\em Chandra} Candidate Counterparts to IGR J22014+6034\label{tab:j22014}}
\begin{minipage}{\linewidth}
\begin{center}
\begin{tabular}{ccccccc} \hline \hline
Source & $\theta$\footnote{The angular distance between the center of the {\em INTEGRAL} error circle, which is also the approximate {\em Chandra} aimpoint, and the source.} & {\em Chandra} R.A. & {\em Chandra} Decl. & ACIS  & Position & \\
Number & (arcminutes) & (J2000)  & (J2000) &  Counts\footnote{The number of ACIS-I counts detected (after background subtraction) in the 0.3--10 keV band.} & Uncertainty\footnote{The 95\% confidence statistical uncertainty on the position using Equation 5 from \cite{hong05}.  This does not include the systematic pointing error, which is $0.\!^{\prime\prime}64$ at 90\% confidence and $1^{\prime\prime}$ at 99\% confidence.} & Hardness\footnote{The hardness is given by $(C_{2}-C_{1})/(C_{2}+C_{1})$, where $C_{2}$ is the number of counts in the 2--10 keV band and $C_{1}$ is the number of counts in the 0.3--2 keV band.}\\ \hline
 1 & 0.97 & $22^{\rm h}01^{\rm m}25^{\rm s}.01$ & +$60^{\circ}34^{\prime}03.\!^{\prime\prime}0$ &  4.4 & $0.\!^{\prime\prime}62$ & ---\\ 
 2 & 1.26 & $22^{\rm h}01^{\rm m}36^{\rm s}.82$ & +$60^{\circ}33^{\prime}50.\!^{\prime\prime}2$ &  3.4 & $0.\!^{\prime\prime}74$ & ---\\ 
 3 & 1.79 & $22^{\rm h}01^{\rm m}15^{\rm s}.12$ & +$60^{\circ}32^{\prime}55.\!^{\prime\prime}2$ &  8.4 & $0.\!^{\prime\prime}56$ &  +$0.56\pm 0.61$\\ 
 4 & 2.13 & $22^{\rm h}01^{\rm m}09^{\rm s}.34$ & +$60^{\circ}34^{\prime}08.\!^{\prime\prime}7$ & 14.4 & $0.\!^{\prime\prime}50$ & --$0.79\pm 0.46$\\ 
 5 & 2.41 & $22^{\rm h}01^{\rm m}46^{\rm s}.26$ & +$60^{\circ}33^{\prime}56.\!^{\prime\prime}9$ &  5.4 & $0.\!^{\prime\prime}78$ & --$0.43\pm 0.82$\\ 
 6 & 3.27 & $22^{\rm h}01^{\rm m}13^{\rm s}.28$ & +$60^{\circ}31^{\prime}11.\!^{\prime\prime}5$ &  4.4 & $1.\!^{\prime\prime}19$ & ---\\ 
 7 & 3.27 & $22^{\rm h}01^{\rm m}26^{\rm s}.91$ & +$60^{\circ}30^{\prime}44.\!^{\prime\prime}7$ &  3.4 & $1.\!^{\prime\prime}41$ & ---\\ 
 8 & 3.44 & $22^{\rm h}01^{\rm m}06^{\rm s}.43$ & +$60^{\circ}31^{\prime}38.\!^{\prime\prime}8$ &  3.4 & $1.\!^{\prime\prime}51$ & ---\\ 
 9 & 3.51 & $22^{\rm h}01^{\rm m}16^{\rm s}.68$ & +$60^{\circ}37^{\prime}18.\!^{\prime\prime}6$ &  8.4 & $0.\!^{\prime\prime}89$ &  +$0.79\pm 0.67$\\ 
10 & 3.87 & $22^{\rm h}01^{\rm m}57^{\rm s}.40$ & +$60^{\circ}33^{\prime}11.\!^{\prime\prime}5$ &  6.4 & $1.\!^{\prime\prime}17$ &  +$0.73\pm 0.79$\\ 
11 & 4.49 & $22^{\rm h}01^{\rm m}30^{\rm s}.94$ & +$60^{\circ}29^{\prime}33.\!^{\prime\prime}6$ &  2.7 & $2.\!^{\prime\prime}82$ & ---\\ 
12 & 4.61 & $22^{\rm h}01^{\rm m}15^{\rm s}.05$ & +$60^{\circ}38^{\prime}24.\!^{\prime\prime}1$ &  4.7 & $1.\!^{\prime\prime}93$ & ---\\ 
13 & 4.67 & $22^{\rm h}01^{\rm m}18^{\rm s}.31$ & +$60^{\circ}38^{\prime}34.\!^{\prime\prime}5$ &  2.7 & $3.\!^{\prime\prime}06$ & ---\\ 
14 & 5.34 & $22^{\rm h}01^{\rm m}54^{\rm s}.13$ & +$60^{\circ}38^{\prime}09.\!^{\prime\prime}5$ & 31.7 & $0.\!^{\prime\prime}81$ & --$0.61\pm 0.27$\\ 
15 & 5.89 & $22^{\rm h}02^{\rm m}13^{\rm s}.69$ & +$60^{\circ}32^{\prime}54.\!^{\prime\prime}4$ &  4.7 & $3.\!^{\prime\prime}25$ & ---\\ 
16 & 6.11 & $22^{\rm h}02^{\rm m}08^{\rm s}.35$ & +$60^{\circ}30^{\prime}42.\!^{\prime\prime}4$ & 10.7 & $1.\!^{\prime\prime}85$ & --$1.15\pm 0.70$\\ 
17 & 6.12 & $22^{\rm h}01^{\rm m}21^{\rm s}.87$ & +$60^{\circ}40^{\prime}06.\!^{\prime\prime}9$ & 28.7 & $1.\!^{\prime\prime}04$ &  +$0.13\pm 0.25$\\ 
18 & 6.13 & $22^{\rm h}02^{\rm m}15^{\rm s}.84$ & +$60^{\circ}35^{\prime}01.\!^{\prime\prime}3$ &  4.7 & $3.\!^{\prime\prime}58$ & ---\\ 
19 & 6.19 & $22^{\rm h}01^{\rm m}30^{\rm s}.48$ & +$60^{\circ}40^{\prime}11.\!^{\prime\prime}3$ &  4.7 & $3.\!^{\prime\prime}67$ & ---\\ 
20 & 6.32 & $22^{\rm h}01^{\rm m}41^{\rm s}.45$ & +$60^{\circ}40^{\prime}04.\!^{\prime\prime}2$ &  2.7 & $6.\!^{\prime\prime}32$ & ---\\ 
21 & 6.94 & $22^{\rm h}00^{\rm m}31^{\rm s}.63$ & +$60^{\circ}32^{\prime}27.\!^{\prime\prime}1$ &  3.8 & $5.\!^{\prime\prime}99$ & ---\\ 
22 & 7.29 & $22^{\rm h}00^{\rm m}58^{\rm s}.94$ & +$60^{\circ}27^{\prime}34.\!^{\prime\prime}7$ & 16.8 & $1.\!^{\prime\prime}98$ &  +$0.19\pm 0.38$\\ 
23 & 7.87 & $22^{\rm h}02^{\rm m}08^{\rm s}.45$ & +$60^{\circ}28^{\prime}04.\!^{\prime\prime}0$ &  2.8 & $11.\!^{\prime\prime}50$ & ---\\ 
24 & 8.05 & $22^{\rm h}00^{\rm m}21^{\rm s}.14$ & +$60^{\circ}33^{\prime}41.\!^{\prime\prime}8$ & 45.8 & $1.\!^{\prime\prime}35$ & --$0.91\pm 0.25$\\ 
25 & 8.90 & $22^{\rm h}02^{\rm m}03^{\rm s}.44$ & +$60^{\circ}26^{\prime}21.\!^{\prime\prime}8$ &  3.8 & $12.\!^{\prime\prime}53$ & ---\\ 
26 & 9.55 & $22^{\rm h}01^{\rm m}43^{\rm s}.90$ & +$60^{\circ}24^{\prime}42.\!^{\prime\prime}7$ &  8.8 & $6.\!^{\prime\prime}80$ & --$1.24\pm 0.91$\\ 
27 & 10.02 & $22^{\rm h}01^{\rm m}40^{\rm s}.46$ & +$60^{\circ}24^{\prime}09.\!^{\prime\prime}0$ & 18.8 & $4.\!^{\prime\prime}00$ &  +$0.17\pm 0.35$\\ 
28 & 10.38 & $22^{\rm h}02^{\rm m}03^{\rm s}.26$ & +$60^{\circ}24^{\prime}40.\!^{\prime\prime}3$ &  3.8 & $20.\!^{\prime\prime}66$ & ---\\ \hline 
\end{tabular}
\end{center}
\end{minipage}
\end{table*}

\end{document}